\documentclass[prl,nofootinbib,preprintnumbers,tightenlines,twocolumn,superscriptaddress]{revtex4-2}
\usepackage{braket}
\usepackage{array}
\usepackage{dsfont}
\usepackage{amsmath}
\usepackage{pifont}
\usepackage[normalem]{ulem}
\usepackage{amssymb}
\usepackage{xcolor}
\usepackage{amsfonts}
\usepackage{mathtools}
\usepackage{dcolumn}
\usepackage{bm}
\usepackage{multirow}
\usepackage{hhline}
\usepackage[pdftex]{graphicx,hyperref}
\usepackage{leftidx}
\usepackage{color}
\usepackage{mathtools}
\usepackage{MnSymbol}
\usepackage[mathscr]{eucal}
\usepackage[german,english]{babel}
\usepackage[capitalize]{cleveref}
\usepackage[utf8]{inputenc}
\usepackage{natbib}
\usepackage{comment}

\begin{document}

\title{Spin-valley entangled quantum Hall states in graphene}
\author{Nikolaos Stefanidis}
\email{nikos@pks.mpg.de}
\affiliation{Max-Planck-Institut für Physik komplexer Systeme, Dresden 01187, Germany}
\author{Inti Sodemann Villadiego}
\email{sodemann@uni-leipzig.de} 
\affiliation{Institut für Theoretische Physik, Universität Leipzig, D-04103, Leipzig, Germany}

\begin{abstract}
We investigate interaction-driven integer quantum Hall states realized in Landau levels of monolayer graphene when two out of its four nearly degenerate spin-valley flavors are filled. By employing a model that accounts for interactions beyond pure delta-functions as well as Zeeman and substrate-induced valley potentials, we demonstrate the existence of a delicate competition of several phases with spontaneous generation of spin-valley entanglement, akin to the spontaneous appearance of spin-orbit coupling driven by interactions. We encounter a particular phase that we term the entangled-Kekul\'{e}-antiferromagnet (E-KD-AF) which only becomes spin-valley entangled under the simultaneous presence of Zeeman and substrate potentials, because it gains energy by simultaneously canting in the spin and valley spaces, by combining features of a canted anti-ferromagnet and a canted Kekul\'{e} state. We quantify the degree of spin-valley entanglement of the many competing phases by computing their bipartite concurrence.

\end{abstract}

\maketitle

\noindent

{\it \textcolor{blue}{Introduction.}} The phase diagram of monolayer graphene in strong magnetic fields continues to present puzzles. At charge neutrality in the $N=0$ Landau level it is still debated whether graphene is in a Canted Antiferromgnet (CAF), as proposed in transport and magnon transmission experiments \cite{young2014tunable,wei2018electrical,zhou2022strong,paul2022electrically,stepanov2018long}, or in a Kekul\'{e} (KD) state as visualized in STM experiments \cite{coissard2022imaging,liu2022visualizing,li2019scanning,farahi2023broken}. In higher Landau levels the nature of states remains much less clear and the experimental evidence much more limited \cite{yang2021experimental}.

\begin{figure}[t!] 
     \centering  
\includegraphics[width=0.48\textwidth]{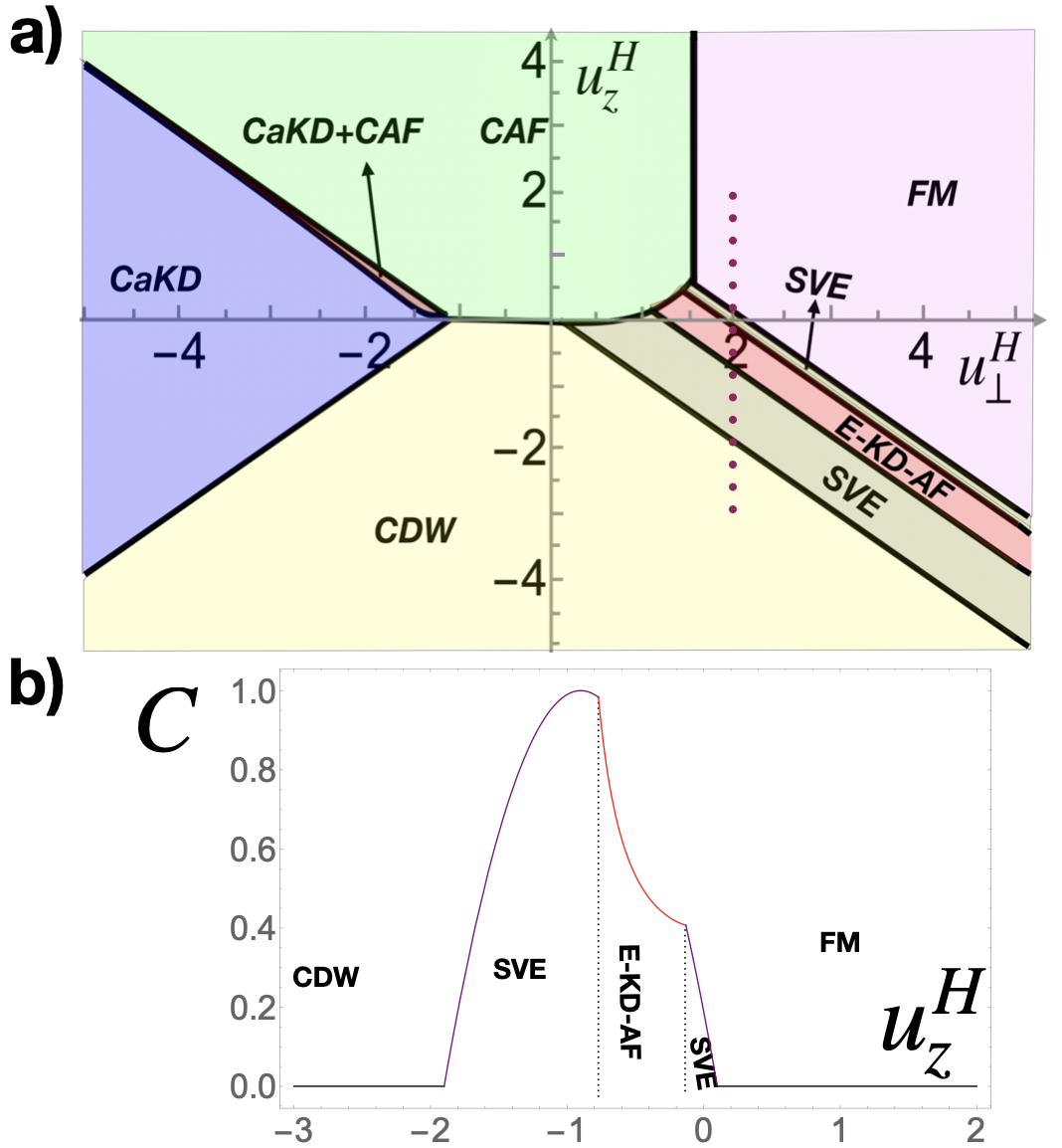}
     \caption{a) Integer quantum Hall states of half-filled Landau levels in graphene with Zeeman, $\epsilon_z=1$, and valley potential, $\epsilon_v=0.1$, and non-delta function interactions with  $\Delta_{\perp}=2, \ \Delta_z=1$ (see Eq.\eqref{uHperp}). The spin-valley entangled state E-KD-AF 
     appears between the two SVE states from Ref.~\cite{de2023global}. b) The concurrence ($C$) measure of spin-valley entanglement is plotted for the cut shown in Fig.~\ref{Fig2}(a) at $u^{H}_{\perp}=2$.}
     \label{Fig2}
\end{figure}

Reference~\cite{kharitonov2012phase} introduced an important model that simplified the understanding of symmetry broken states relative to earlier studies \cite{nomura2006quantum,goerbig2006electron,alicea2006graphene,herbut2007theory,jung2009theory} by capturing the valley symmetry breaking interactions in the $N=0$ Landau level as pure delta function interactions. Recent studies, however, have emphasized the need to consider interactions beyond delta functions  in higher Landau levels~\cite{yang2021experimental,stefanidis2023competing}, and also in the $N=0$ Landau level arising from Landau level mixing~\cite{das2022coexistence,de2023global}. In this work we investigate the interplay of such longer range interactions with the presence of spin Zeeman and substrate-induced sub-lattice symmetry breaking potentials, within a model that is applicable to integer quantum Hall states of graphene in any of its Landau levels. We will demonstrate that the combination of these ingredients leads to an interesting competition of phases with spontaneous spin-valley entanglement. Interestingly we find a state which becomes entangled only under the simultaneous presence of spin and valley Zeeman terms and interactions with longer range than pure delta functions, which we term the Entangled-Kekul\'{e}-Antiferromagnet state (E-KD-AF) (see Fig.\ref{Fig2}).



\begin{figure*}[t!] 
     \centering  
     \includegraphics[width=0.95\textwidth]{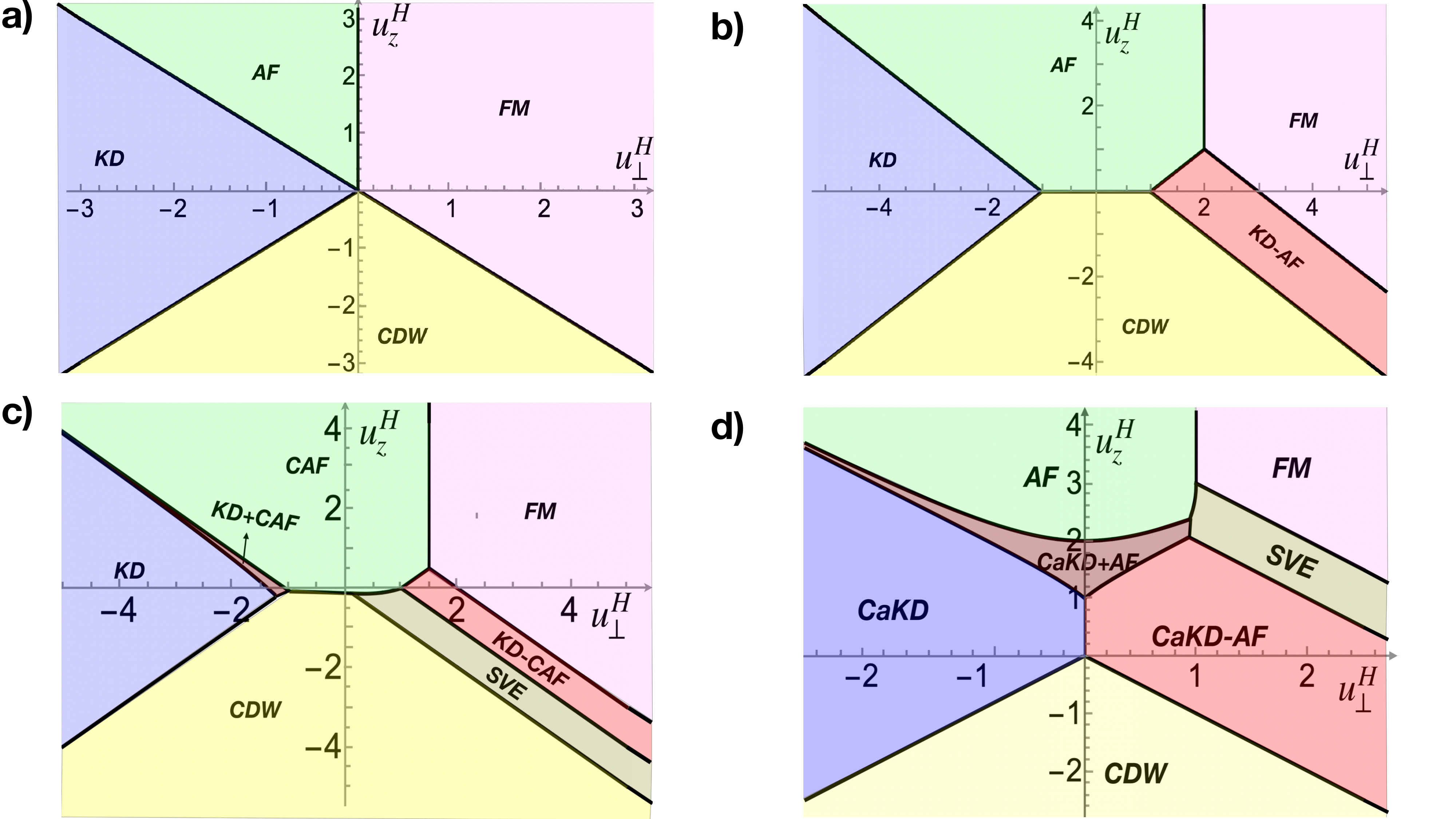}
     \caption{Phase diagrams of integer quantum Hall states in graphene for half-filled Landau levels: a) delta-function interactions and $\epsilon_z=\epsilon_v=0$ from Ref.~\cite{kharitonov2012phase}. b) For interactions with finite range, $\Delta_z =1, \ \Delta_{\perp}=2$,  and $\epsilon_z=\epsilon_v=0$, showing that the KD-AF becomes favourable. c) For interactions with finite range, $\Delta_z =1, \ \Delta_{\perp}=2$, and $\epsilon_z=1, \ \epsilon_v=0$, showing that the KD-AF cants and becomes the KD-CAF and the SVE phase~\cite{de2023global} becomes favourable. d) For interactions with finite range, $\Delta_z =2, \ \Delta_{\perp}=1$, and $\epsilon_v=1, \ \epsilon_z=0$ showing that the  KD-AF cants in the valleys space and becomes the CaKD-AF and the SVE phase of Ref.~\cite{de2023global} appears in between the KD-AF and the FM states.}
     \label{Fig1}
\end{figure*}

\noindent

{\it \textcolor{blue}{Model, mean-field theory, and entanglement measure.}} 
A series of recent works have considered the following continuum model of the projected interaction Hamiltonian onto the N-th Landau level of graphene~\cite{stefanidis2023competing,das2022coexistence,de2023global,atteia2021skyrmion}:
\begin{equation} \label{proj model}
\mathcal{H}^{N}=\sum_{i<j} \{ V^{N}_{z}(r_{ij}) \tau^{i}_z\tau^{j}_z + V^{N} _{\perp}(r_{ij})\tau^{i}_{\perp}\tau^{j}_{\perp}\}-\epsilon_Z\sum_{i}s^{i}_{z}-\epsilon_V\sum_{i}\tau^{i}_{z},
\end{equation}
where $V^N_{z,\perp}(r_{ij})$ are interactions that depend only on distance $r_{ij}$ between particles $i,j$, $\tau_{\perp}^i \tau_{\perp}^j=\tau_x^i \tau_x^j+\tau_y^i \tau_y^j$ and $s_a, 
\tau_{a}, \ a=0,...,3$ are the Pauli matrices acting on the valleys and spin respectively. This model captures the symmetry breaking terms beyond the $SU(4)$ invariant long-range part of the Coulomb interaction. This model goes beyond the model of Ref.~\cite{kharitonov2012phase} which can be viewed as a limit of Eq.~\ref{proj model} when the interactions become delta functions, $V_{z, \perp}(r_{ij})=V_{z, \perp} \delta(r_{i}-r_{j})$. Refs.~\cite{young2014tunable,stefanidis2023competing} have demonstrated that even for models of unprojected interactions that are short-ranged (see e.g. Ref.~\cite{aleiner2007spontaneous}), effective interactions will naturally appear as a result of the projection onto higher Landau levels ($N\neq 0$). It has been also recently emphasized that corrections to pure delta functions appear naturally in higher Landau levels~\cite{stefanidis2023competing} by projecting the general model of short-distance interactions of graphene of Ref.~\cite{aleiner2007spontaneous}, but can also appear even in the $N=0$ LL due to Landau level mixing effects~\cite{das2022coexistence,de2023global}.
\par When there is an integer-filling of Landau levels, the Hartree-Fock variational energy functional of translationally invariant quantum Hall ferromagnets for the above model can be written as~\cite{stefanidis2023competing}:
\begin{equation} \label{HF energy}
\begin{split}
\mathcal{E}_{HF}[P] &= \frac{1}{2} \sum_{i=x,y,z}\bigg( u^{H}_{i} (Tr \{ \tau_{i} P\})^2-u^{X}_{i}Tr \{ (\tau_{i} P)^2\} \bigg) \\
&-\epsilon_z Tr \{ s_z P\}- \epsilon_v Tr \{ \eta_z P\}, \\
\end{split}
\end{equation}
here $P$ is the projector into the occupied spinors, which in the case of half-filling (two filled components) equals $P=\ket{F_1} \bra{F_1}+\ket{F_2} \bra{F_2}$, where $\ket{F_i}, \ i=1,2$  are arbitrary orthonormal vectors within the four-dimensional Hilbert space of spin and valley flavors. The HF energy function is parametrized by four independent interaction energy scales $u_z^H,u_z^X,u_x^{H,X}=u_y^{H,X}=u_{\perp}^{H,X}$, given by:

\begin{equation}\label{uHperp}
u^{H}_{a}=\frac{V_{a}(\mathbf{q}=0)}{8 \pi^2},
    u^{X}_{a}=\frac{1}{8 \pi^2}\iint d \mathbf{q}V_{a}(\mathbf{q}), \ a=\perp,z.
\end{equation}


\par In the limit of pure delta function interactions, the difference between Hartree and exchange energy constants, $\Delta_{z, \perp}=u^{H}_{z, \perp}-u^{X}_{z, \perp}$ would vanish, and we would have only two interaction constants, as in the model of Ref. ~\cite{kharitonov2012phase}. We will consider general spin-valley entangled variational states~\cite{ezawa2005ground,atteia2021skyrmion,das2022coexistence,lian2017spin, atteia2021skyrmion, stefanidis2023competing, de2023global}:
\begin{equation} \label{entangled states}
\begin{split}
\ket{F}_1&=\cos\frac{a_1}{2} \ket{\boldsymbol{\eta}}\ket{\mathbf{s}}+e^{i \beta_1} \sin \frac{a_1}{2} \ket{-\boldsymbol{\eta}} \ket{-\mathbf{s}}, \\
\ket{F}_2&=\cos\frac{a_2}{2} \ket{\boldsymbol{\eta}}\ket{\mathbf{-s}}+e^{i \beta_2} \sin \frac{a_2}{2} \ket{-\boldsymbol{\eta}} \ket{\mathbf{s}}.\\
\end{split}
\end{equation} 
Here  $\ket{\boldsymbol{\eta}}$ and $\ket{\mathbf{s}}$ are states parametrized by unit vectors $\boldsymbol{\eta}$ and $\mathbf{s}$ in the spin and valley Bloch spheres respectively and $a_{1,2}$ and $\beta_{1,2}$ are real constants. Because the projector $P$ is effectively a mixed state, simple measures of bipartite entanglement applicable to pure states, such as the von-Neumann entropy of the reduced density matrix, are not suitable. Instead, the degree of spin-valley bipartite entanglement associated with the projector $P$ onto the above two states, can be measured by the concurrence $C$ defined as~\cite{knothe2015edge,mintert2005measures}:
\begin{equation}
    C\equiv {\rm Max} \{\lambda_1-\lambda_2-\lambda_3-\lambda_4,0\},
\end{equation}
where $\lambda_i$ are the eigenvalues of the matrix $R= P (\tau_{y} \bigotimes s_{y}) P^{T} (\tau_{y} \bigotimes s_{y}) P$, ordered according to $\lambda_i \geq \lambda_j$, for $i>j$. For projector onto the states in Eq.~\eqref{entangled states}, the concurrence is:
\begin{equation}\label{formula entanglement}
    C=|\sin ^2 a_{1}-\sin ^2 a_{2}|.
\end{equation}
\begin{table*}[h!t!]
\begin{tabular}{ |p{7cm}||p{8cm}||p{2.5cm}|} 
 \hline
 \multicolumn{3}{|c|}{States appearing for two filled Landau levels} \\
 \hline
States & Coordinates  & Concurrence $C$ \\
 \hline
 CDW (Charge density wave)  & $a_1=a_2=\theta_p=0$ & $0$\\
CaKD (Canted Kekul\'{e} distortion)& $a_1=a_2=0,\  \theta_p=\cos^{-1} \big( \frac{\epsilon_v}{u^{H}_z-u^{H}_{\perp}+(\Delta_z-\Delta_{\perp})} \big)$ & $0$ \\
 FM (Ferromagnet)& $a_1=0, \ a_2=\pi,\  \theta_p=\theta_s=0$& $0$  \\
 CAF  (Canted Antiferromagnet)  & $a_1=a, \ \cos a=-\frac{\epsilon_z}{2u^{X}_{\perp}}, \ a_2 = \pi-a, \ \theta_p=\pi/2, \ \theta_s=0, \ \beta=0$ & $0$ \\
 E-KD-AF (Entangled Kekul\'{e} Distortion antiferromagnet)& $a_{1,2}=\cos^{-1} \{\pm \frac{\epsilon_z}{\Delta_{\perp}+\Delta_z-u^{H}_{\perp}-u^{H}_z}+\frac{\epsilon_v}{-\Delta_{\perp}+\Delta_z+u^{H}_{\perp}+u^{H}_z}\}, \ \theta_p=0, \ \theta_s=0, \ \beta=0$ & $|\sin^2 a_1-\sin^2 a_2|$ \\
  SVE (Spin-valley entangled phase of Ref.~\cite{de2023global})&$a_1=0,  \ \cos a_2 = \frac{\epsilon_z-\epsilon_v+u^{H}_z+u^{H}_{\perp}-\Delta_{\perp}}{-\Delta_z}, \ \theta_p=0, \ \theta_s=0, \ \beta=0$ & $\sin^2 a_2$\\
 \hline
\end{tabular}
\caption{Competing states and their coordinates. The KD and KD-CAF are obtained by taking the limit of $\epsilon_v \rightarrow 0$ from the CaKD and CaKD-CAF respectively while the AF and CaKD-AF by taking the limit of $\epsilon_z \rightarrow 0$ from the CAF and CaKD-CAF respectively. In the $a_{1,2}$ coordinates of E-KD-AF the $+$ ($-$) sign corresponds to $a_{1}$($a_{2}$).}
\label{Table1}
\end{table*}
When the minima of HF energy are spin-valley disentangled states, we have $C=0$, and these states can be separated into two classes, one of ``valley-active'' states with spinors given by:

\begin{equation}\label{valleyactive}    \ket{F}_1=\ket{\boldsymbol{\eta}_1}\ket{\mathbf{s}}, \ \ket{F}_2=\ket{\boldsymbol{\eta}_2}\ket{-\mathbf{s}}, 
\end{equation}


\noindent where $\boldsymbol{\eta}_1,\boldsymbol{\eta}_2$ are two arbirtary directions in the valley Bloch sphere, and another class of ``spin active'' states, with spinors given by:

\begin{equation}
\ket{F}_1=\ket{\boldsymbol{\eta}}\ket{\mathbf{s}_1}, \ \ket{F}_2=\ket{\boldsymbol{-\eta}}\ket{\mathbf{s}_2}.
\end{equation}

\noindent where $\boldsymbol{s}_1,\boldsymbol{s}_2$ are two arbirtary directions in the valley Bloch sphere.

\par In the limit of pure delta function interactions ($\Delta_{z,\perp}=0$), Ref.~\cite{kharitonov2012phase} found a phase diagram with four spin-valley disentangled states that we reproduce in Fig~\ref{Fig1}.(a): FM (Ferromagnet), AF (Antiferromagnet), KD (Kekul\'{e} distortion), and CDW (Charge density wave). When interactions are not pure delta functions, and in the absence of Zeeman and valley potentials ($\epsilon_z=\epsilon_v=0$), we recently found in Ref.~\cite{stefanidis2023competing} that a new phase termed the KD-AF (Kekul\'{e}- Antiferromagnet) can appear, as shown in Fig.~\ref{Fig2}(a). However in the absence of Zeeman and valley potentials ($\epsilon_z=\epsilon_v=0$) all these five states have no spin-valley quantum entanglement.
In particular, the KD-AF phase can be viewed as one of valley-active states from Eq.\eqref{valleyactive} having one component occupying an equal amplitude superposition of both valleys (e.g. $\boldsymbol{\eta}_1=\boldsymbol{\hat{x}}$) with one spin and the other component occupying the opposite valley coherent superposition (e.g. $\boldsymbol{\eta}_2=-\boldsymbol{\hat{x}}$) with the opposite spin, and therefore has a non-trivial spin-valley correlation, but no spin-valley entanglement properly speaking.


In this work, we will show that these five states (FM, AF, KD, CDW, KD-AF) can be viewed as parent states to several spin-valley entangled phases. Some of them, such as the KD/AF coexistence and SVE states identified in Refs.~\cite{de2023global,das2022coexistence}, arise near the phase boundaries between these parent states after adding Zeeman and valley potentials. However, we will also show that among these five parent states the KD-AF is special because it is the only one that becomes spin-valley entangled under the simultaneous presence of Zeeman and valley sublattice potentials, and we will term the state that evolves continuously from the KD-AF under these perturbations the entangled-Kekul\'{e}-antiferromagnet (E-KD-AF) state (see Appendix S-VI for details of comparison with Ref.~\cite{de2023global}).

\noindent

{\it \textcolor{blue}{Ground states with either Zeeman or valley potentials.}} We begin our analysis by studying the phase diagram when only the Zeeman coupling, $\epsilon_z \neq 0, \ \epsilon_v=0$, is present in Eq.~\eqref{proj model}. We find that both the AF and the KD-AF cant their spins, which is a natural tendency of the anti-ferromagnetic states
in order to take advantage of the Zeeman energy, evolving into the CAF and KD-CAF states depicted in Fig.~\ref{Fig2}c). These two states remain, however, spin-valley disentangled. The KD-CAF appears in between the FM and the CDW as long as $0<\epsilon_z<2\Delta_z$. 
\par However, as pointed out in Refs.~\cite{das2022coexistence,de2023global}, the CAF and the KD become unstable over some region close to their boundary leading to a mixed state of AFM-Kekul\'{e} phase coexistence which occupies a thin sliver of the phase boundary between these two phases. The analytic coordinates for this coexistence state are discussed in Appendix S-III. This phase coexistence occurs only when $\Delta_{\perp}>0$ and otherwise there is a direct first order phase transition between the Kekul\'{e} and CAF states. Additionally, a finite $\epsilon_z$ induces the formation of another a new phase, the SVE of Ref.~\cite{de2023global} growing from the boundary of the CDW with the KD-CAF. For $\epsilon_z=\epsilon_v=0$ the SVE phase is never the ground state over any finite region, but interestingly it is is degenerate with KD-AF only at its boundaries with the FM and CDW. We note that the degeneracy at the boundary with the FM persists for all values of the Zeeman field, making this boundary presumably of higher symmetry~\cite{wu2014so}. When $\epsilon_z >0$ and $ \epsilon_v=0$, the SVE, therefore, starts nucleating at the boundary of the CDW and the KD-AF and grows with increasing $\epsilon_z$ until it occupies the whole region between the CDW and the FM at a critical value of the Zeeman field $\epsilon^{c}_z=2 \Delta_z$. The transition of KD-CAF with the FM is continuous, i.e the spin of the KD-CAF cants continuously until it reaches the fully polarized value of $s_{z}=2$. The KD-CAF is therefore expected to have the similar signatures as the standard CAF state in spin sensitive probes, such as the magnon transmission experiments of Ref.~\cite{young2014tunable}.

\par It is also useful to consider the limit when ($\epsilon_v \neq 0$) is present but the Zeeman coupling vanishes ($\epsilon_z=0$). This leads to the canting of the KD, similarly to the $N=0$ Landau level, as discussed in Refs.~\cite{hegde2022theory} and also as shown in Fig.~\ref{Fig1}(d). Interestingly, since the KD-AF is simultaneously anti-ferromagnetic in the valley space and in the spin space, it will undergo canting of the valley pseudo-spins towards the z-axis driven by the finite $\epsilon_v$. We also find that the $\epsilon_v$ also induces an intermediate coexistence region at the boundary between CaKD and AF analogous to the coexistence region of Ref.~\cite{das2022coexistence,de2023global}. (see Fig.~\ref{Fig1}d). On the other hand, for $\epsilon_v>0$ and $\epsilon_z=0$, the SVE state now starts growing from the boundary of the FM with the KD-AF whereas the SVE is always degenerate with the KD-AF at its boundary with the CDW. The CaKD-AF persists until a critical value of the valley Zeeman, $\epsilon^{c}_v=2 \Delta_z$.

\noindent

{\it \textcolor{blue}{Ground states with both Zeeman and valley potentials.}} We now turn to the general case where both the Zeeman coupling and the hBN substrate are present. Our results are illustrated in Fig.~\ref{Fig2}a). We again find a coexistence of the CaKD and the CAF along a sliver of the phase diagram. However, the main qualitative difference is that the KD-AF state transforms into a new spin-valley entangled state that we call the E-KD-AF when both spin and valley Zeeman fields are simultaneously present, as depicted in Fig.~\ref{Fig2}b). This tendency orginates from the fact that the KD-AF state gains energy by canting either in the spin and valley direction under the presence of spin or valley Zeeman terms, but it is impossible to construct disentangled states that cant simultaneously in this way (see Table~\ref{Table1}). We have found the exact coordinates of the spin-valley entangled minima of the Hartree-Fock functional in Eq.~\eqref{HF energy} and they satisfy $\beta=\theta_s=\theta_p=0$, which is shared by all the phases in the right two quadrants of the phase diagrams. The E-KD-AF is now sandwiched between two spin-valley entangled SVE phases of Ref.~\cite{de2023global} in the region between the FM and the CDW (see Fig.~\ref{Fig2}a)), yet represents a qualitative distinct phase. 

\par We can distinguish the two competing spin-valley entangled phases, namely the E-KD-AFM and the SVE of Ref.~\cite{de2023global} by their order parameters, $\hat{O}_{ij}=Tr\{P \tau_i s_j\}$ (see supplementary material for further details). Both of them have a vanishing total of valley and spin in the $x-y$ plane, $\hat{O}_{a0}=\hat{O}_{0a}=0$, with $a=x,y$. However, the SVE phase of Ref.~\cite{de2023global}, has the order parameters $\hat{O}_{xx},  \hat{O}_{yy}$ locked to be equal $\hat{O}_{xx}=\hat{O}_{yy}=\sin a$, while for the E-KD-AF these order parameters are generally distinct and given by $\hat{O}_{xx}=\sin a_1 +\sin a_2, \hat{O}_{yy}=-\sin a_1 +\sin a_2$ (see Table~\ref{Table1} for the values of $a_{1,2}$). Moreover, as illustrated in Fig.~\ref{Fig2}b), the concurrence of the SVE is different than that of the E-KD-AF (see also Table~\ref{Table1}).

{\it \textcolor{blue}{Summary and Discussion.}}
We have studied the integer quantum Hall ferromagnet states of graphene within a model applicable to any of its Landau levels, and focused on the case of half-filling when two out of four of its nearly degenerate spin-valley states are filled. Our model accounts for valley symmetry-breaking interactions beyond pure delta functions, and  includes the simulatneous presence of the Zeeman coupling and a substrate-induced valley symmetry breaking potential (e.g. from alignment with a hBN substrate). We have computed the concurrence measure of entanglement which allows us to quantify the degree of spin-valley entanglement of these states.

\par Besides the known spin-valley disentangled states such as the antiferromagnet and the Kekul\'{e} valence-bond-solid, we have found a delicate competition of states featuring spontaneous spin-valley entanglement, akin to that arising from spin-orbit coupling, but whose origin stems purely from interaction driven spontaneous symmetry breaking. Notably, we have found a state which only becomes entangled under the simultaneous presence of spin and valley Zeeman terms and interactions with longer range than pure delta functions, which we term the Entangled-Kekul\'{e}-Antiferromagnet state (E-KD-AF). This tendency arises because this state combines features of the anti-ferromagnet and the Kekul\'{e} states, and the state tries to cant simultaneously in the spin and valley Bloch sphere in order to gain energy from these single particle terms, but it can only achieve this at the expense of becoming spin-valley entangled.



{\it \textcolor{blue}{Acknowledgements.}} We would like to thank Ganpathy Murthy, Chunli Huang and Nemin Wei for valuable discussions. NS acknowledges useful discussions with Panagiotis Giannakeas and Hongzheng Zhao. We acknowledge
support by the Deutsche Forschungsgemeinschaft (DFG) through a research grant with project number 518372354.

\bibliography{ref}

\begin{thebibliography}{27}%
\makeatletter
\providecommand \@ifxundefined [1]{%
 \@ifx{#1\undefined}
}%
\providecommand \@ifnum [1]{%
 \ifnum #1\expandafter \@firstoftwo
 \else \expandafter \@secondoftwo
 \fi
}%
\providecommand \@ifx [1]{%
 \ifx #1\expandafter \@firstoftwo
 \else \expandafter \@secondoftwo
 \fi
}%
\providecommand \natexlab [1]{#1}%
\providecommand \enquote  [1]{``#1''}%
\providecommand \bibnamefont  [1]{#1}%
\providecommand \bibfnamefont [1]{#1}%
\providecommand \citenamefont [1]{#1}%
\providecommand \href@noop [0]{\@secondoftwo}%
\providecommand \href [0]{\begingroup \@sanitize@url \@href}%
\providecommand \@href[1]{\@@startlink{#1}\@@href}%
\providecommand \@@href[1]{\endgroup#1\@@endlink}%
\providecommand \@sanitize@url [0]{\catcode `\\12\catcode `\$12\catcode
  `\&12\catcode `\#12\catcode `\^12\catcode `\_12\catcode `\%12\relax}%
\providecommand \@@startlink[1]{}%
\providecommand \@@endlink[0]{}%
\providecommand \url  [0]{\begingroup\@sanitize@url \@url }%
\providecommand \@url [1]{\endgroup\@href {#1}{\urlprefix }}%
\providecommand \urlprefix  [0]{URL }%
\providecommand \Eprint [0]{\href }%
\providecommand \doibase [0]{https://doi.org/}%
\providecommand \selectlanguage [0]{\@gobble}%
\providecommand \bibinfo  [0]{\@secondoftwo}%
\providecommand \bibfield  [0]{\@secondoftwo}%
\providecommand \translation [1]{[#1]}%
\providecommand \BibitemOpen [0]{}%
\providecommand \bibitemStop [0]{}%
\providecommand \bibitemNoStop [0]{.\EOS\space}%
\providecommand \EOS [0]{\spacefactor3000\relax}%
\providecommand \BibitemShut  [1]{\csname bibitem#1\endcsname}%
\let\auto@bib@innerbib\@empty
\bibitem [{\citenamefont {Young}\ \emph {et~al.}(2014)\citenamefont {Young},
  \citenamefont {Sanchez-Yamagishi}, \citenamefont {Hunt}, \citenamefont
  {Choi}, \citenamefont {Watanabe}, \citenamefont {Taniguchi}, \citenamefont
  {Ashoori},\ and\ \citenamefont {Jarillo-Herrero}}]{young2014tunable}%
  \BibitemOpen
  \bibfield  {author} {\bibinfo {author} {\bibfnamefont {A.}~\bibnamefont
  {Young}}, \bibinfo {author} {\bibfnamefont {J.}~\bibnamefont
  {Sanchez-Yamagishi}}, \bibinfo {author} {\bibfnamefont {B.}~\bibnamefont
  {Hunt}}, \bibinfo {author} {\bibfnamefont {S.}~\bibnamefont {Choi}}, \bibinfo
  {author} {\bibfnamefont {K.}~\bibnamefont {Watanabe}}, \bibinfo {author}
  {\bibfnamefont {T.}~\bibnamefont {Taniguchi}}, \bibinfo {author}
  {\bibfnamefont {R.}~\bibnamefont {Ashoori}},\ and\ \bibinfo {author}
  {\bibfnamefont {P.}~\bibnamefont {Jarillo-Herrero}},\ }\bibfield  {title}
  {\bibinfo {title} {Tunable symmetry breaking and helical edge transport in a
  graphene quantum spin hall state},\ }\href@noop {} {\bibfield  {journal}
  {\bibinfo  {journal} {Nature}\ }\textbf {\bibinfo {volume} {505}},\ \bibinfo
  {pages} {528} (\bibinfo {year} {2014})}\BibitemShut {NoStop}%
\bibitem [{\citenamefont {Wei}\ \emph {et~al.}(2018)\citenamefont {Wei},
  \citenamefont {Van Der~Sar}, \citenamefont {Lee}, \citenamefont {Watanabe},
  \citenamefont {Taniguchi}, \citenamefont {Halperin},\ and\ \citenamefont
  {Yacoby}}]{wei2018electrical}%
  \BibitemOpen
  \bibfield  {author} {\bibinfo {author} {\bibfnamefont {D.~S.}\ \bibnamefont
  {Wei}}, \bibinfo {author} {\bibfnamefont {T.}~\bibnamefont {Van Der~Sar}},
  \bibinfo {author} {\bibfnamefont {S.~H.}\ \bibnamefont {Lee}}, \bibinfo
  {author} {\bibfnamefont {K.}~\bibnamefont {Watanabe}}, \bibinfo {author}
  {\bibfnamefont {T.}~\bibnamefont {Taniguchi}}, \bibinfo {author}
  {\bibfnamefont {B.~I.}\ \bibnamefont {Halperin}},\ and\ \bibinfo {author}
  {\bibfnamefont {A.}~\bibnamefont {Yacoby}},\ }\bibfield  {title} {\bibinfo
  {title} {Electrical generation and detection of spin waves in a quantum hall
  ferromagnet},\ }\href@noop {} {\bibfield  {journal} {\bibinfo  {journal}
  {Science}\ }\textbf {\bibinfo {volume} {362}},\ \bibinfo {pages} {229}
  (\bibinfo {year} {2018})}\BibitemShut {NoStop}%
\bibitem [{\citenamefont {Zhou}\ \emph {et~al.}(2022)\citenamefont {Zhou},
  \citenamefont {Huang}, \citenamefont {Wei}, \citenamefont {Taniguchi},
  \citenamefont {Watanabe}, \citenamefont {Zaletel}, \citenamefont {Papi{\'c}},
  \citenamefont {MacDonald},\ and\ \citenamefont {Young}}]{zhou2022strong}%
  \BibitemOpen
  \bibfield  {author} {\bibinfo {author} {\bibfnamefont {H.}~\bibnamefont
  {Zhou}}, \bibinfo {author} {\bibfnamefont {C.}~\bibnamefont {Huang}},
  \bibinfo {author} {\bibfnamefont {N.}~\bibnamefont {Wei}}, \bibinfo {author}
  {\bibfnamefont {T.}~\bibnamefont {Taniguchi}}, \bibinfo {author}
  {\bibfnamefont {K.}~\bibnamefont {Watanabe}}, \bibinfo {author}
  {\bibfnamefont {M.~P.}\ \bibnamefont {Zaletel}}, \bibinfo {author}
  {\bibfnamefont {Z.}~\bibnamefont {Papi{\'c}}}, \bibinfo {author}
  {\bibfnamefont {A.~H.}\ \bibnamefont {MacDonald}},\ and\ \bibinfo {author}
  {\bibfnamefont {A.~F.}\ \bibnamefont {Young}},\ }\bibfield  {title} {\bibinfo
  {title} {Strong-magnetic-field magnon transport in monolayer graphene},\
  }\href@noop {} {\bibfield  {journal} {\bibinfo  {journal} {Physical Review
  X}\ }\textbf {\bibinfo {volume} {12}},\ \bibinfo {pages} {021060} (\bibinfo
  {year} {2022})}\BibitemShut {NoStop}%
\bibitem [{\citenamefont {Paul}\ \emph {et~al.}(2022)\citenamefont {Paul},
  \citenamefont {Sahu}, \citenamefont {Watanabe}, \citenamefont {Taniguchi},
  \citenamefont {Jain}, \citenamefont {Murthy},\ and\ \citenamefont
  {Das}}]{paul2022electrically}%
  \BibitemOpen
  \bibfield  {author} {\bibinfo {author} {\bibfnamefont {A.~K.}\ \bibnamefont
  {Paul}}, \bibinfo {author} {\bibfnamefont {M.~R.}\ \bibnamefont {Sahu}},
  \bibinfo {author} {\bibfnamefont {K.}~\bibnamefont {Watanabe}}, \bibinfo
  {author} {\bibfnamefont {T.}~\bibnamefont {Taniguchi}}, \bibinfo {author}
  {\bibfnamefont {J.}~\bibnamefont {Jain}}, \bibinfo {author} {\bibfnamefont
  {G.}~\bibnamefont {Murthy}},\ and\ \bibinfo {author} {\bibfnamefont
  {A.}~\bibnamefont {Das}},\ }\bibfield  {title} {\bibinfo {title}
  {Electrically switchable tunneling across a graphene pn junction: evidence
  for canted antiferromagnetic phase in $\nu=0$ state},\ }\href@noop {}
  {\bibfield  {journal} {\bibinfo  {journal} {arXiv preprint arXiv:2205.00710}\
  } (\bibinfo {year} {2022})}\BibitemShut {NoStop}%
\bibitem [{\citenamefont {Stepanov}\ \emph {et~al.}(2018)\citenamefont
  {Stepanov}, \citenamefont {Che}, \citenamefont {Shcherbakov}, \citenamefont
  {Yang}, \citenamefont {Chen}, \citenamefont {Thilahar}, \citenamefont
  {Voigt}, \citenamefont {Bockrath}, \citenamefont {Smirnov}, \citenamefont
  {Watanabe} \emph {et~al.}}]{stepanov2018long}%
  \BibitemOpen
  \bibfield  {author} {\bibinfo {author} {\bibfnamefont {P.}~\bibnamefont
  {Stepanov}}, \bibinfo {author} {\bibfnamefont {S.}~\bibnamefont {Che}},
  \bibinfo {author} {\bibfnamefont {D.}~\bibnamefont {Shcherbakov}}, \bibinfo
  {author} {\bibfnamefont {J.}~\bibnamefont {Yang}}, \bibinfo {author}
  {\bibfnamefont {R.}~\bibnamefont {Chen}}, \bibinfo {author} {\bibfnamefont
  {K.}~\bibnamefont {Thilahar}}, \bibinfo {author} {\bibfnamefont
  {G.}~\bibnamefont {Voigt}}, \bibinfo {author} {\bibfnamefont {M.~W.}\
  \bibnamefont {Bockrath}}, \bibinfo {author} {\bibfnamefont {D.}~\bibnamefont
  {Smirnov}}, \bibinfo {author} {\bibfnamefont {K.}~\bibnamefont {Watanabe}},
  \emph {et~al.},\ }\bibfield  {title} {\bibinfo {title} {Long-distance spin
  transport through a graphene quantum hall antiferromagnet},\ }\href@noop {}
  {\bibfield  {journal} {\bibinfo  {journal} {Nature Physics}\ }\textbf
  {\bibinfo {volume} {14}},\ \bibinfo {pages} {907} (\bibinfo {year}
  {2018})}\BibitemShut {NoStop}%
\bibitem [{\citenamefont {Coissard}\ \emph {et~al.}(2022)\citenamefont
  {Coissard}, \citenamefont {Wander}, \citenamefont {Vignaud}, \citenamefont
  {Grushin}, \citenamefont {Repellin}, \citenamefont {Watanabe}, \citenamefont
  {Taniguchi}, \citenamefont {Gay}, \citenamefont {Winkelmann}, \citenamefont
  {Courtois} \emph {et~al.}}]{coissard2022imaging}%
  \BibitemOpen
  \bibfield  {author} {\bibinfo {author} {\bibfnamefont {A.}~\bibnamefont
  {Coissard}}, \bibinfo {author} {\bibfnamefont {D.}~\bibnamefont {Wander}},
  \bibinfo {author} {\bibfnamefont {H.}~\bibnamefont {Vignaud}}, \bibinfo
  {author} {\bibfnamefont {A.~G.}\ \bibnamefont {Grushin}}, \bibinfo {author}
  {\bibfnamefont {C.}~\bibnamefont {Repellin}}, \bibinfo {author}
  {\bibfnamefont {K.}~\bibnamefont {Watanabe}}, \bibinfo {author}
  {\bibfnamefont {T.}~\bibnamefont {Taniguchi}}, \bibinfo {author}
  {\bibfnamefont {F.}~\bibnamefont {Gay}}, \bibinfo {author} {\bibfnamefont
  {C.~B.}\ \bibnamefont {Winkelmann}}, \bibinfo {author} {\bibfnamefont
  {H.}~\bibnamefont {Courtois}}, \emph {et~al.},\ }\bibfield  {title} {\bibinfo
  {title} {Imaging tunable quantum hall broken-symmetry orders in graphene},\
  }\href@noop {} {\bibfield  {journal} {\bibinfo  {journal} {Nature}\ }\textbf
  {\bibinfo {volume} {605}},\ \bibinfo {pages} {51} (\bibinfo {year}
  {2022})}\BibitemShut {NoStop}%
\bibitem [{\citenamefont {Liu}\ \emph {et~al.}(2022)\citenamefont {Liu},
  \citenamefont {Farahi}, \citenamefont {Chiu}, \citenamefont {Papic},
  \citenamefont {Watanabe}, \citenamefont {Taniguchi}, \citenamefont
  {Zaletel},\ and\ \citenamefont {Yazdani}}]{liu2022visualizing}%
  \BibitemOpen
  \bibfield  {author} {\bibinfo {author} {\bibfnamefont {X.}~\bibnamefont
  {Liu}}, \bibinfo {author} {\bibfnamefont {G.}~\bibnamefont {Farahi}},
  \bibinfo {author} {\bibfnamefont {C.-L.}\ \bibnamefont {Chiu}}, \bibinfo
  {author} {\bibfnamefont {Z.}~\bibnamefont {Papic}}, \bibinfo {author}
  {\bibfnamefont {K.}~\bibnamefont {Watanabe}}, \bibinfo {author}
  {\bibfnamefont {T.}~\bibnamefont {Taniguchi}}, \bibinfo {author}
  {\bibfnamefont {M.~P.}\ \bibnamefont {Zaletel}},\ and\ \bibinfo {author}
  {\bibfnamefont {A.}~\bibnamefont {Yazdani}},\ }\bibfield  {title} {\bibinfo
  {title} {Visualizing broken symmetry and topological defects in a quantum
  hall ferromagnet},\ }\href@noop {} {\bibfield  {journal} {\bibinfo  {journal}
  {Science}\ }\textbf {\bibinfo {volume} {375}},\ \bibinfo {pages} {321}
  (\bibinfo {year} {2022})}\BibitemShut {NoStop}%
\bibitem [{\citenamefont {Li}\ \emph {et~al.}(2019)\citenamefont {Li},
  \citenamefont {Zhang}, \citenamefont {Yin},\ and\ \citenamefont
  {He}}]{li2019scanning}%
  \BibitemOpen
  \bibfield  {author} {\bibinfo {author} {\bibfnamefont {S.-Y.}\ \bibnamefont
  {Li}}, \bibinfo {author} {\bibfnamefont {Y.}~\bibnamefont {Zhang}}, \bibinfo
  {author} {\bibfnamefont {L.-J.}\ \bibnamefont {Yin}},\ and\ \bibinfo {author}
  {\bibfnamefont {L.}~\bibnamefont {He}},\ }\bibfield  {title} {\bibinfo
  {title} {Scanning tunneling microscope study of quantum hall isospin
  ferromagnetic states in the zero landau level in a graphene monolayer},\
  }\href@noop {} {\bibfield  {journal} {\bibinfo  {journal} {Physical Review
  B}\ }\textbf {\bibinfo {volume} {100}},\ \bibinfo {pages} {085437} (\bibinfo
  {year} {2019})}\BibitemShut {NoStop}%
\bibitem [{\citenamefont {Farahi}\ \emph {et~al.}(2023)\citenamefont {Farahi},
  \citenamefont {Chiu}, \citenamefont {Liu}, \citenamefont {Papic},
  \citenamefont {Watanabe}, \citenamefont {Taniguchi}, \citenamefont
  {Zaletel},\ and\ \citenamefont {Yazdani}}]{farahi2023broken}%
  \BibitemOpen
  \bibfield  {author} {\bibinfo {author} {\bibfnamefont {G.}~\bibnamefont
  {Farahi}}, \bibinfo {author} {\bibfnamefont {C.-L.}\ \bibnamefont {Chiu}},
  \bibinfo {author} {\bibfnamefont {X.}~\bibnamefont {Liu}}, \bibinfo {author}
  {\bibfnamefont {Z.}~\bibnamefont {Papic}}, \bibinfo {author} {\bibfnamefont
  {K.}~\bibnamefont {Watanabe}}, \bibinfo {author} {\bibfnamefont
  {T.}~\bibnamefont {Taniguchi}}, \bibinfo {author} {\bibfnamefont {M.~P.}\
  \bibnamefont {Zaletel}},\ and\ \bibinfo {author} {\bibfnamefont
  {A.}~\bibnamefont {Yazdani}},\ }\bibfield  {title} {\bibinfo {title} {Broken
  symmetries and excitation spectra of interacting electrons in partially
  filled landau levels},\ }\href@noop {} {\bibfield  {journal} {\bibinfo
  {journal} {arXiv preprint arXiv:2303.16993}\ } (\bibinfo {year}
  {2023})}\BibitemShut {NoStop}%
\bibitem [{\citenamefont {Yang}\ \emph {et~al.}(2021)\citenamefont {Yang},
  \citenamefont {Zibrov}, \citenamefont {Bai}, \citenamefont {Taniguchi},
  \citenamefont {Watanabe}, \citenamefont {Zaletel},\ and\ \citenamefont
  {Young}}]{yang2021experimental}%
  \BibitemOpen
  \bibfield  {author} {\bibinfo {author} {\bibfnamefont {F.}~\bibnamefont
  {Yang}}, \bibinfo {author} {\bibfnamefont {A.~A.}\ \bibnamefont {Zibrov}},
  \bibinfo {author} {\bibfnamefont {R.}~\bibnamefont {Bai}}, \bibinfo {author}
  {\bibfnamefont {T.}~\bibnamefont {Taniguchi}}, \bibinfo {author}
  {\bibfnamefont {K.}~\bibnamefont {Watanabe}}, \bibinfo {author}
  {\bibfnamefont {M.~P.}\ \bibnamefont {Zaletel}},\ and\ \bibinfo {author}
  {\bibfnamefont {A.~F.}\ \bibnamefont {Young}},\ }\bibfield  {title} {\bibinfo
  {title} {Experimental determination of the energy per particle in partially
  filled landau levels},\ }\href@noop {} {\bibfield  {journal} {\bibinfo
  {journal} {Physical review letters}\ }\textbf {\bibinfo {volume} {126}},\
  \bibinfo {pages} {156802} (\bibinfo {year} {2021})}\BibitemShut {NoStop}%
\bibitem [{\citenamefont {Kharitonov}(2012)}]{kharitonov2012phase}%
  \BibitemOpen
  \bibfield  {author} {\bibinfo {author} {\bibfnamefont {M.}~\bibnamefont
  {Kharitonov}},\ }\bibfield  {title} {\bibinfo {title} {Phase diagram for the
  {$\nu= 0$} quantum hall state in monolayer graphene},\ }\href@noop {}
  {\bibfield  {journal} {\bibinfo  {journal} {Physical Review B}\ }\textbf
  {\bibinfo {volume} {85}},\ \bibinfo {pages} {155439} (\bibinfo {year}
  {2012})}\BibitemShut {NoStop}%
\bibitem [{\citenamefont {Nomura}\ and\ \citenamefont
  {MacDonald}(2006)}]{nomura2006quantum}%
  \BibitemOpen
  \bibfield  {author} {\bibinfo {author} {\bibfnamefont {K.}~\bibnamefont
  {Nomura}}\ and\ \bibinfo {author} {\bibfnamefont {A.~H.}\ \bibnamefont
  {MacDonald}},\ }\bibfield  {title} {\bibinfo {title} {Quantum hall
  ferromagnetism in graphene},\ }\href@noop {} {\bibfield  {journal} {\bibinfo
  {journal} {Physical review letters}\ }\textbf {\bibinfo {volume} {96}},\
  \bibinfo {pages} {256602} (\bibinfo {year} {2006})}\BibitemShut {NoStop}%
\bibitem [{\citenamefont {Goerbig}\ \emph {et~al.}(2006)\citenamefont
  {Goerbig}, \citenamefont {Moessner},\ and\ \citenamefont
  {Dou{\c{c}}ot}}]{goerbig2006electron}%
  \BibitemOpen
  \bibfield  {author} {\bibinfo {author} {\bibfnamefont {M.~O.}\ \bibnamefont
  {Goerbig}}, \bibinfo {author} {\bibfnamefont {R.}~\bibnamefont {Moessner}},\
  and\ \bibinfo {author} {\bibfnamefont {B.}~\bibnamefont {Dou{\c{c}}ot}},\
  }\bibfield  {title} {\bibinfo {title} {Electron interactions in graphene in a
  strong magnetic field},\ }\href@noop {} {\bibfield  {journal} {\bibinfo
  {journal} {Physical Review B}\ }\textbf {\bibinfo {volume} {74}},\ \bibinfo
  {pages} {161407} (\bibinfo {year} {2006})}\BibitemShut {NoStop}%
\bibitem [{\citenamefont {Alicea}\ and\ \citenamefont
  {Fisher}(2006)}]{alicea2006graphene}%
  \BibitemOpen
  \bibfield  {author} {\bibinfo {author} {\bibfnamefont {J.}~\bibnamefont
  {Alicea}}\ and\ \bibinfo {author} {\bibfnamefont {M.~P.}\ \bibnamefont
  {Fisher}},\ }\bibfield  {title} {\bibinfo {title} {Graphene integer quantum
  hall effect in the ferromagnetic and paramagnetic regimes},\ }\href@noop {}
  {\bibfield  {journal} {\bibinfo  {journal} {Physical Review B}\ }\textbf
  {\bibinfo {volume} {74}},\ \bibinfo {pages} {075422} (\bibinfo {year}
  {2006})}\BibitemShut {NoStop}%
\bibitem [{\citenamefont {Herbut}(2007)}]{herbut2007theory}%
  \BibitemOpen
  \bibfield  {author} {\bibinfo {author} {\bibfnamefont {I.~F.}\ \bibnamefont
  {Herbut}},\ }\bibfield  {title} {\bibinfo {title} {Theory of integer quantum
  hall effect in graphene},\ }\href@noop {} {\bibfield  {journal} {\bibinfo
  {journal} {Physical Review B}\ }\textbf {\bibinfo {volume} {75}},\ \bibinfo
  {pages} {165411} (\bibinfo {year} {2007})}\BibitemShut {NoStop}%
\bibitem [{\citenamefont {Jung}\ and\ \citenamefont
  {MacDonald}(2009)}]{jung2009theory}%
  \BibitemOpen
  \bibfield  {author} {\bibinfo {author} {\bibfnamefont {J.}~\bibnamefont
  {Jung}}\ and\ \bibinfo {author} {\bibfnamefont {A.}~\bibnamefont
  {MacDonald}},\ }\bibfield  {title} {\bibinfo {title} {Theory of the
  magnetic-field-induced insulator in neutral graphene sheets},\ }\href@noop {}
  {\bibfield  {journal} {\bibinfo  {journal} {Physical Review B}\ }\textbf
  {\bibinfo {volume} {80}},\ \bibinfo {pages} {235417} (\bibinfo {year}
  {2009})}\BibitemShut {NoStop}%
\bibitem [{\citenamefont {Stefanidis}\ and\ \citenamefont
  {Villadiego}(2023)}]{stefanidis2023competing}%
  \BibitemOpen
  \bibfield  {author} {\bibinfo {author} {\bibfnamefont {N.}~\bibnamefont
  {Stefanidis}}\ and\ \bibinfo {author} {\bibfnamefont {I.~S.}\ \bibnamefont
  {Villadiego}},\ }\bibfield  {title} {\bibinfo {title} {Competing spin-valley
  entangled and broken symmetry states in the n= 1 landau level of graphene},\
  }\href@noop {} {\bibfield  {journal} {\bibinfo  {journal} {Physical Review
  B}\ }\textbf {\bibinfo {volume} {107}},\ \bibinfo {pages} {045132} (\bibinfo
  {year} {2023})}\BibitemShut {NoStop}%
\bibitem [{\citenamefont {Das}\ \emph {et~al.}(2022)\citenamefont {Das},
  \citenamefont {Kaul},\ and\ \citenamefont {Murthy}}]{das2022coexistence}%
  \BibitemOpen
  \bibfield  {author} {\bibinfo {author} {\bibfnamefont {A.}~\bibnamefont
  {Das}}, \bibinfo {author} {\bibfnamefont {R.~K.}\ \bibnamefont {Kaul}},\ and\
  \bibinfo {author} {\bibfnamefont {G.}~\bibnamefont {Murthy}},\ }\bibfield
  {title} {\bibinfo {title} {Coexistence of canted antiferromagnetism and bond
  order in $\nu$= 0 graphene},\ }\href@noop {} {\bibfield  {journal} {\bibinfo
  {journal} {Physical Review Letters}\ }\textbf {\bibinfo {volume} {128}},\
  \bibinfo {pages} {106803} (\bibinfo {year} {2022})}\BibitemShut {NoStop}%
\bibitem [{\citenamefont {De}\ \emph {et~al.}(2023)\citenamefont {De},
  \citenamefont {Das}, \citenamefont {Rao}, \citenamefont {Kaul},\ and\
  \citenamefont {Murthy}}]{de2023global}%
  \BibitemOpen
  \bibfield  {author} {\bibinfo {author} {\bibfnamefont {S.~J.}\ \bibnamefont
  {De}}, \bibinfo {author} {\bibfnamefont {A.}~\bibnamefont {Das}}, \bibinfo
  {author} {\bibfnamefont {S.}~\bibnamefont {Rao}}, \bibinfo {author}
  {\bibfnamefont {R.~K.}\ \bibnamefont {Kaul}},\ and\ \bibinfo {author}
  {\bibfnamefont {G.}~\bibnamefont {Murthy}},\ }\bibfield  {title} {\bibinfo
  {title} {Global phase diagram of charge-neutral graphene in the quantum hall
  regime for generic interactions},\ }\href@noop {} {\bibfield  {journal}
  {\bibinfo  {journal} {Physical Review B}\ }\textbf {\bibinfo {volume}
  {107}},\ \bibinfo {pages} {125422} (\bibinfo {year} {2023})}\BibitemShut
  {NoStop}%
\bibitem [{\citenamefont {Atteia}\ \emph {et~al.}(2021)\citenamefont {Atteia},
  \citenamefont {Lian},\ and\ \citenamefont {Goerbig}}]{atteia2021skyrmion}%
  \BibitemOpen
  \bibfield  {author} {\bibinfo {author} {\bibfnamefont {J.}~\bibnamefont
  {Atteia}}, \bibinfo {author} {\bibfnamefont {Y.}~\bibnamefont {Lian}},\ and\
  \bibinfo {author} {\bibfnamefont {M.~O.}\ \bibnamefont {Goerbig}},\
  }\bibfield  {title} {\bibinfo {title} {Skyrmion zoo in graphene at charge
  neutrality in a strong magnetic field},\ }\href@noop {} {\bibfield  {journal}
  {\bibinfo  {journal} {Physical Review B}\ }\textbf {\bibinfo {volume}
  {103}},\ \bibinfo {pages} {035403} (\bibinfo {year} {2021})}\BibitemShut
  {NoStop}%
\bibitem [{\citenamefont {Aleiner}\ \emph {et~al.}(2007)\citenamefont
  {Aleiner}, \citenamefont {Kharzeev},\ and\ \citenamefont
  {Tsvelik}}]{aleiner2007spontaneous}%
  \BibitemOpen
  \bibfield  {author} {\bibinfo {author} {\bibfnamefont {I.~L.}\ \bibnamefont
  {Aleiner}}, \bibinfo {author} {\bibfnamefont {D.~E.}\ \bibnamefont
  {Kharzeev}},\ and\ \bibinfo {author} {\bibfnamefont {A.~M.}\ \bibnamefont
  {Tsvelik}},\ }\bibfield  {title} {\bibinfo {title} {Spontaneous symmetry
  breaking in graphene subjected to an in-plane magnetic field},\ }\href@noop
  {} {\bibfield  {journal} {\bibinfo  {journal} {Physical Review B}\ }\textbf
  {\bibinfo {volume} {76}},\ \bibinfo {pages} {195415} (\bibinfo {year}
  {2007})}\BibitemShut {NoStop}%
\bibitem [{\citenamefont {Ezawa}\ \emph {et~al.}(2005)\citenamefont {Ezawa},
  \citenamefont {Eliashvili},\ and\ \citenamefont
  {Tsitsishvili}}]{ezawa2005ground}%
  \BibitemOpen
  \bibfield  {author} {\bibinfo {author} {\bibfnamefont {Z.}~\bibnamefont
  {Ezawa}}, \bibinfo {author} {\bibfnamefont {M.}~\bibnamefont {Eliashvili}},\
  and\ \bibinfo {author} {\bibfnamefont {G.}~\bibnamefont {Tsitsishvili}},\
  }\bibfield  {title} {\bibinfo {title} {Ground-state structure in $\nu$= 2
  bilayer quantum hall systems},\ }\href@noop {} {\bibfield  {journal}
  {\bibinfo  {journal} {Physical Review B}\ }\textbf {\bibinfo {volume} {71}},\
  \bibinfo {pages} {125318} (\bibinfo {year} {2005})}\BibitemShut {NoStop}%
\bibitem [{\citenamefont {Lian}\ and\ \citenamefont
  {Goerbig}(2017)}]{lian2017spin}%
  \BibitemOpen
  \bibfield  {author} {\bibinfo {author} {\bibfnamefont {Y.}~\bibnamefont
  {Lian}}\ and\ \bibinfo {author} {\bibfnamefont {M.~O.}\ \bibnamefont
  {Goerbig}},\ }\bibfield  {title} {\bibinfo {title} {Spin-valley skyrmions in
  graphene at filling factor {$\nu$}=- 1},\ }\href@noop {} {\bibfield
  {journal} {\bibinfo  {journal} {Physical Review B}\ }\textbf {\bibinfo
  {volume} {95}},\ \bibinfo {pages} {245428} (\bibinfo {year}
  {2017})}\BibitemShut {NoStop}%
\bibitem [{\citenamefont {Knothe}\ and\ \citenamefont
  {Jolicoeur}(2015)}]{knothe2015edge}%
  \BibitemOpen
  \bibfield  {author} {\bibinfo {author} {\bibfnamefont {A.}~\bibnamefont
  {Knothe}}\ and\ \bibinfo {author} {\bibfnamefont {T.}~\bibnamefont
  {Jolicoeur}},\ }\bibfield  {title} {\bibinfo {title} {Edge structure of
  graphene monolayers in the $\nu$= 0 quantum hall state},\ }\href@noop {}
  {\bibfield  {journal} {\bibinfo  {journal} {Physical Review B}\ }\textbf
  {\bibinfo {volume} {92}},\ \bibinfo {pages} {165110} (\bibinfo {year}
  {2015})}\BibitemShut {NoStop}%
\bibitem [{\citenamefont {Mintert}\ \emph {et~al.}(2005)\citenamefont
  {Mintert}, \citenamefont {Carvalho}, \citenamefont {Ku{\'s}},\ and\
  \citenamefont {Buchleitner}}]{mintert2005measures}%
  \BibitemOpen
  \bibfield  {author} {\bibinfo {author} {\bibfnamefont {F.}~\bibnamefont
  {Mintert}}, \bibinfo {author} {\bibfnamefont {A.~R.}\ \bibnamefont
  {Carvalho}}, \bibinfo {author} {\bibfnamefont {M.}~\bibnamefont {Ku{\'s}}},\
  and\ \bibinfo {author} {\bibfnamefont {A.}~\bibnamefont {Buchleitner}},\
  }\bibfield  {title} {\bibinfo {title} {Measures and dynamics of entangled
  states},\ }\href@noop {} {\bibfield  {journal} {\bibinfo  {journal} {Physics
  Reports}\ }\textbf {\bibinfo {volume} {415}},\ \bibinfo {pages} {207}
  (\bibinfo {year} {2005})}\BibitemShut {NoStop}%
\bibitem [{\citenamefont {Wu}\ \emph {et~al.}(2014)\citenamefont {Wu},
  \citenamefont {Sodemann}, \citenamefont {Araki}, \citenamefont {MacDonald},\
  and\ \citenamefont {Jolicoeur}}]{wu2014so}%
  \BibitemOpen
  \bibfield  {author} {\bibinfo {author} {\bibfnamefont {F.}~\bibnamefont
  {Wu}}, \bibinfo {author} {\bibfnamefont {I.}~\bibnamefont {Sodemann}},
  \bibinfo {author} {\bibfnamefont {Y.}~\bibnamefont {Araki}}, \bibinfo
  {author} {\bibfnamefont {A.~H.}\ \bibnamefont {MacDonald}},\ and\ \bibinfo
  {author} {\bibfnamefont {T.}~\bibnamefont {Jolicoeur}},\ }\bibfield  {title}
  {\bibinfo {title} {So (5) symmetry in the quantum hall effect in graphene},\
  }\href@noop {} {\bibfield  {journal} {\bibinfo  {journal} {Physical Review
  B}\ }\textbf {\bibinfo {volume} {90}},\ \bibinfo {pages} {235432} (\bibinfo
  {year} {2014})}\BibitemShut {NoStop}%
\bibitem [{\citenamefont {Hegde}\ and\ \citenamefont
  {Villadiego}(2022)}]{hegde2022theory}%
  \BibitemOpen
  \bibfield  {author} {\bibinfo {author} {\bibfnamefont {S.~S.}\ \bibnamefont
  {Hegde}}\ and\ \bibinfo {author} {\bibfnamefont {I.~S.}\ \bibnamefont
  {Villadiego}},\ }\bibfield  {title} {\bibinfo {title} {Theory of competing
  charge density wave, kekul{\'e}, and antiferromagnetically ordered fractional
  quantum hall states in graphene aligned with boron nitride},\ }\href@noop {}
  {\bibfield  {journal} {\bibinfo  {journal} {Physical Review B}\ }\textbf
  {\bibinfo {volume} {105}},\ \bibinfo {pages} {195417} (\bibinfo {year}
  {2022})}\BibitemShut {NoStop}%
\end{thebibliography}%

\clearpage

\renewcommand{\theequation}{S-\arabic{equation}}
\renewcommand{\thefigure}{S-\arabic{figure}}
\renewcommand{\thetable}{S-\Roman{table}}
\makeatletter
\renewcommand\@biblabel[1]{S#1.}
\setcounter{equation}{0}
\setcounter{figure}{0}

\onecolumngrid

\begin{center}
\textbf{\large Supplemental Material: Spin-valley entangled quantum Hall states in graphene}
\end{center}

\section{S-I: Brief review of the model and Hartree-Fock theory}
In Ref.~\cite{stefanidis2023competing}, we introduced a model for the higher Landau levels of graphene which arises from projecting all the possible valley dependent short range interactions. This is based on the model of Ref.~\cite{aleiner2007spontaneous}, which describes the short-range electron-electron interactions in the absence of a magnetic field and can be written in the following way: 
\begin{equation} \label{S-Asymmetry}
\mathcal{H}_A= \sum_{i<j}\{ \sum_{\alpha,\beta}V_{\alpha \beta} T^{i}_{\alpha \beta} T^{j}_{\alpha \beta}\} \delta(\mathbf{r}_i-\mathbf{r}_j),
\end{equation}
where we have defined $T^{i(j)}_{\alpha \beta}=\tau_{\alpha}^{i(j)} \otimes \sigma_{\beta}^{i(j)}\otimes s_{0}^{i(j)}$, and $\tau_{\alpha}^{i(j)}, \ \sigma_{\beta}^{i(j)}, \ s_{0}^{i(j)}$ $\alpha,\beta=0,x,y,z$ to be the Pauli matrices acting on valley, sublattice and spin respectively. The lattice symmetries of graphene introduce certain constraints on the interaction strengths~\cite{aleiner2007spontaneous}:
\begin{equation} \label{S-symmetries}
\begin{split}
&F_{\perp z}  \equiv V_{xx}=V_{yx}, \\
&F_{z \perp}\equiv V_{0x}=V_{zy}, \\
&F_{\perp \perp}\equiv V_{xz}=V_{y0}=V_{yz}=V_{x0}, \\
&F_{0 \perp}\equiv V_{zx}=V_{0y}, \\
&F_{\perp 0}\equiv V_{yy}=V_{xy}, \\
&F_{zz}\equiv V_{0z}, \\
&F_{z0}\equiv V_{z0}, \\
&F_{0z}\equiv V_{zz}, \\
\end{split}
\end{equation}
yielding only $9$ real independent parameters. The projected model in the higher Landau levels takes the form~\cite{stefanidis2023competing}:
\begin{equation} \label{S-proj model}
\mathcal{H}^{N}_{A}=\sum_{i<j} \{ V^{N}_{z}(r_{ij}) \tau^{i}_z\tau^{j}_z + V^{N} _{\perp}(r_{ij})\tau^{i}_{\perp}\tau^{j}_{\perp}\},
\end{equation}
with $\tau^{i}_{\perp}\tau^{j}_{\perp}=\tau^{i}_{x}\tau^{j}_{x}+\tau^{i}_{y}\tau^{j}_{y}$ and the strengths now containing longer range interactions:
\begin{equation} \label{S-strengths N=1}
        V_{z,\perp}(r_{ij}) = \sum_{n=0}^{2}g^{z,\perp}_n \nabla^{2n} \delta(r_{ij}).\\
\end{equation}
The model of Eq.~\eqref{S-proj model} would also contain non-delta function effective interactions arising from Landau level mixing~\cite{das2022coexistence,de2023global}, and therefore by taking $V_z$, and $V_{\perp}$ to be arbitrary functions, we can also view it as a general model for any Landau level of graphene. Then, the Hartree-Fock functional for translational invariant quantum hall ferromagnets, including the Zeeman effect and hBN substrate, reads:
\begin{equation} \label{S-General HFS}
\begin{split}
&\mathcal{E}_{HF}= \frac{1}{2} \bigg\{ 2u^{H}_{z} \big( M_{z}^{1} M_{z}^{2}- |M_{z}^{12}|^2\big) +2u^{H}_{\perp} \big( M_{x}^{1} M_{x}^{2}+ M_{y}^{1} M_{y}^{2}- |M_{x}^{12}|^2- |M_{y}^{12}|^2\big) \\
&+\Delta_z \big( |M_{z}^{1}|^2 + |M_{z}^{2}|^2+ 2 |M_{z}^{12}|^2\big) +\Delta_{\perp} \big( |M_{x}^{1}|^2 + |M_{y}^{2}|^2+ 2 |M_{x}^{12}|^2 +|M_{y}^{1}|^2 + |M_{y}^{2}|^2+ 2 |M_{y}^{12}|^2\big) \bigg \}  \\
&-\epsilon_z s_z- \epsilon_v \eta_{z}, \\\\
\end{split}
\end{equation}
with $M_{a}^{i(j)}=\braket{F_i|\tau_a|F_j}$, where the vectors $F_i$ are defined in Eq.~\eqref{entangled states} of the main text, and we can write the unit vector $\boldsymbol{\eta}$ as ,
$\boldsymbol{\eta}= \begin{pmatrix}
  \sin \theta_p \cos \phi_p \\ 
  \sin \theta_p \sin \phi_p  \\
  \cos \theta_p 
\end{pmatrix}$. We first minimize the Hartree-Fock functional in the valley-spin disentangled spaces. The valley active subspace consists of $\ket{F}_1= \ket{\boldsymbol{\eta}_1}\ket{\mathbf{s}}, \  \ket{F}= \ket{\boldsymbol{\eta}_{2}}\ket{-\mathbf{s}}$:
\begin{equation} \label{S-HF valley}
\begin{split}
\mathcal{E}_{HF} &=\frac{1}{2} \Big\{ \Delta_z (n^2_{1z}+n^2_{1z}) +2 u^{H}_{z} n_{1z} n_{1z}+ \Delta_{\perp} (n^2_{1 \perp}+n^2_{1 \perp}) +2 u^{H}_{\perp} n_{1 \perp} n_{1 \perp}-2 \epsilon_v (n_{1z}+n_{1z}) \Big\},\\
\end{split}
\end{equation}
and of the spin active states , $\ket{F}_1= \ket{\boldsymbol{\eta}}\ket{\mathbf{s}_1}, \  \ket{F}_2= \ket{-\boldsymbol{\eta}}\ket{\mathbf{s}_2}$:
\begin{equation} \label{S-HF spin}
\begin{split}
\mathcal{E}_{HF}&= \frac{1}{2} \Big \{ -(2 u^{X}_{\perp}+u^{X}_{z})(1+ \mathbf{s}_1 \cdot \mathbf{s}_2) - (u^{X}_{\perp} n_{\perp}^2+u^{X}_{z}n_z^2)(1- \mathbf{s}_1 \cdot \mathbf{s}_2) \Big \}- \epsilon_z(s_{1z}+s_{2z}). \\
\end{split}
\end{equation}

\section{S-II: Spin and Valley Canted states in the $N=1$ Landau level}
The minimization without the presence of spin and valley Zeeman was carried out in Ref.~\cite{stefanidis2023competing}. The antiferromagnets, both in spin and valley space, in the presence of Zeeman coupling and hBN substrate respectively, will cant in the $x-y$ plane to take advantage of the spin (valley) Zeeman. 
The spin antiferromagnets i.e the KD-AF and the AF will cant their spins in the $x-y$ plane while the magnitudes of spin remains fixed $|\mathbf{s}_1|=|\mathbf{s}_2|=1$. Moreover, the $U(1)$ symmetry in the $x-y$ plane gives:
\begin{equation} \label{Canted}
\mathbf{s}_{1,2}=(\pm \sqrt{1-s^2_z},0,s_z).
\end{equation}
The energies of the spin active states are : 
\begin{center}
\begin{tabular}{||c c c c||} 
 \hline
 Spin active states & Energies  \\ [0.5ex] 
 \hline
FM & $\mathcal{E}_{FM}=-\frac{1}{2} \{ (2u^{X}_{\perp}+u^{X}_{z})+4\epsilon_z$\} \\ 
 \hline
AF & $\mathcal{E}_{AF}=-u^{X}_{z}$  \\
 \hline
KD-AF & $\mathcal{E}_{KD-AF}=-u^{X}_{\perp}$  \\ [1ex] 
 \hline
\end{tabular}
\end{center}
Now let us consider how each sector $n_z=1$ and $n_{\perp}=1$ behaves due to the Zeeman.
For $n_z=1$:
\begin{equation}
\mathcal{E}_{HF}^{n_z=1}=-(u^{X}_{\perp}+u^{X}_z)- u^{X}_{\perp}\mathbf{s}_1\mathbf{s}_2-\epsilon_z (s_{1z}+s_{2z}).
\end{equation}
Now due to canting ~\eqref{Canted}, the CAF will have : 
\begin{equation}
\begin{split}
\mathcal{E}^{CAF}&=-u^{X}_{z}+\frac{ \epsilon^2_z}{ 2 u^{X}_{\perp}} =-(u^{H}_z-\Delta_z)+\frac{ \epsilon^2_z}{2(u^{H}_z-\Delta_z)},\\
s^{*}_z&=-\frac{\epsilon_z}{2 u^{X}_{\perp}},
\end{split}
\end{equation}
with $\cos a= -\frac{\epsilon_z}{2u^{X}_{z}}$. We note that :
\begin{equation}
s^{*}_z=1 \rightarrow u^{H}_{\perp}=\Delta_{\perp}-\frac{\epsilon_z}{2},
\end{equation}
which coincides with the CAF-FM boundary.
For $n_{\perp}=1$:
\begin{equation}
\mathcal{E}_{HF}^{n_{\perp}=1}=-(3u^{X}_{\perp}+u^{X}_z)- (u^{X}_{\perp}+u^{X}_z)\mathbf{s}_a \mathbf{s}_b-2\epsilon_z (s_{az}+s_{bz}).
\end{equation}
Now due to canting~\eqref{Canted}, the KD-CAF will have : 
\begin{equation}
\begin{split}
\mathcal{E}^{KD-CAF}&=-2u^{X}_{\perp}+\frac{ 2\epsilon^2_z}{u^{X}_{\perp}+u^{X}_z} =-2(u^{H}_{\perp}-\Delta_{\perp})+\frac{ 2\epsilon^2_z}{ (u^{H}_z+u^{H}_{\perp})-(\Delta_z+\Delta_{\perp})}\\
s^{*}_z&=-\frac{\epsilon_z}{ (u^{X}_{\perp}+u^{X}_z)},
\end{split}
\end{equation}
We note that :
\begin{equation}
s^{*}_z=1 \rightarrow u^{H}_{z}=-u^{H}_{\perp}+(\Delta_{\perp}+\Delta_{z})-\epsilon_z,
\end{equation}
which coincides with the KD-CAF-FM boundary.
Similarly for the valley active states due to Eq.~\eqref{S-HF valley}, we find that the CaKD has :
\begin{equation}
\begin{split}
\eta^{{CaKD}}_z&=\frac{\epsilon_v}{u^{H}_z-u^{H}_{\perp}+(\Delta_z-\Delta_{\perp})}, \\
\mathcal{E}_{CaKD}&=2\{ \Delta_{\perp}+u^{H}_{\perp}- \frac{\epsilon^2_v}{u^{H}_z-u^{H}_{\perp}+(\Delta_z-\Delta_{\perp})} \},
\end{split}
\end{equation}
and the CaKD-AF : 
\begin{equation}
\begin{split}
\eta^{{CaKD-AF}}_z&=\frac{\epsilon_v}{u^{H}_{z}+u^{H}_{\perp}+(\Delta_z-\Delta_{\perp})} \\
\mathcal{E}_{CaKD-AF}&=2\{ \Delta_{\perp}-u^{H}_{\perp}-\frac{\epsilon^2_v}{u^{H}_z+u^{H}_{\perp}+(\Delta_z-\Delta_{\perp})} \}.
\end{split}
\end{equation}
\section{S-III: Linear Stability analysis}
By expanding the complete Hartree-Fock functional, $E_{HF} \{a_1,a_2,\theta_p,\theta_s,\beta, \phi_p \}$, in Eq.~\eqref{S-General HFS}we are able to see if the phases which minimize the Hartree-Fock functional in the spin-valley disentangled spaces, remain stable to spin-valley entangled fluctuations. Below we present the instability lines which are defined as the lines where one of the eigenvalues of the bilinear matrix of stability vanishes.
\subsubsection{CDW}
The coordinates of the CDW are $a_1=a_2=\theta_p=0$ . The instability lines and the directions at which they become unstable:
\begin{equation}
\begin{split}
&a_{1,2}: \ u^{H}_z=-u^{H}_{\perp}+(\Delta_{\perp}-\Delta_z) \pm \epsilon_v + \epsilon_v ,\\
&\theta_p: u^{H}_{z}=u^{H}_{\perp}+(\Delta_{\perp}-\Delta_z)+ \epsilon_v.
\end{split}
\end{equation}
\subsubsection{CaKD}
The coordinates of the CaKD are $a_1=a_2=0,\  \theta_p=\cos^{-1} \big( \frac{\epsilon_v}{u^{H}_z-u^{H}_{\perp}+(\Delta_z-\Delta_{\perp})} \big)$. The instability lines and the directions at which they become unstable:
\begin{equation}
\begin{split}
&a_{1,2}: \ 2 \Delta_{\perp}-2 (\Delta_z+5 u^{H}_{\perp}+3 u^{H}_{z})+8 \epsilon_v \cos a_{0}+2 (3 \Delta_{\perp}-3 \Delta_{z}+u^{H}_{\perp}-u^{H}_{z})\cos 2a_0 \\
& \pm \sqrt{64 \epsilon^2_Z+6(\Delta_{\perp}-\Delta_z-u^{H}_{\perp}+u^{H}_{z})^2+2 (\Delta_{\perp}-\Delta_z-u^{H}_{\perp}+u^{H}_{z})^2(-4 \cos 2a_0+\cos 4a_0)}=0\\
&\theta_p: \ \ \  u^{H}_{z}=u^{H}_{\perp}+(\Delta_{\perp}-\Delta_z), \\
& \ \ \ \ \ \ \ \ u^{H}_{z}=u^{H}_{\perp}+(\Delta_{\perp}-\Delta_z) \pm \epsilon_v .
\end{split}
\end{equation}

\subsubsection{FM}
The coordinates of the CaKD are $a_1=0, \ a_2=\pi,\  \theta_p=\theta_s=0$ . The instability lines and the directions at which they become unstable:

\begin{equation}
\begin{split}
&a_{1,2}: \ u^{H}_z=-u^{H}_{\perp}+(\Delta_{\perp}+\Delta_z) \pm \epsilon_v-\epsilon_z .\\
\end{split}
\end{equation}
\subsubsection{CAF}
The coordinates of the CAF are $a_1=0, \ a_2=\pi,\  \theta_p=\theta_s=0$. Defining $\cos a_0=-\frac{\epsilon_z}{2(u^{H}_{\perp}-\Delta_{\perp})}$, the instability lines are :
\begin{equation}
\begin{split}
&a_{1,2}, \theta_p: \ \epsilon_v= \sqrt{(-\Delta_{\perp}+\Delta_z+u_{\perp}-u_z)(\Delta_z-u_{\perp}-u_z-\epsilon_z \cos a_0+\Delta_{\perp} \cos 2a_0)} ,\\
&\ \ \  u_{\perp}=\Delta_{\perp} \pm \frac{\epsilon_z}{2} .\\
\end{split}
\end{equation}
\subsubsection{CaKD-AF}
The coordinates of the CaKD-AF are  $\alpha_{1,2}=\cos^{-1} \{\epsilon_v/(-\Delta_{\perp}+\Delta_z+u^{H}_{\perp}+u^{H}_z)\}, \ \theta_p=0, \ \theta_s=0, \ \beta=0$ . Defining $\cos a_0=\epsilon_v/(u_{z}+u_{\perp}+\Delta_z-\Delta_{\perp})$, the Hartree-Fock functional is : 
\begin{equation}
\begin{split}
E_{HF}&= \frac{1}{2} a^2_1 \{\Delta_{\perp}-u^{H}_{\perp}-u^{H}_{z}+2(\epsilon_v+ \epsilon_z )\cos a_0+(\Delta_{\perp}-2 \Delta_z- u^{H}_{\perp}-u_z) \cos 2a_{0} \} \\
&+\frac{1}{2} a^2_2\{\Delta_{\perp}-u^{H}_{\perp}-u^{H}_{z}+2(\epsilon_v- \epsilon_z )\cos a_0+(\Delta_{\perp}-2 \Delta_z- u^{H}_{\perp}-u_z) \cos 2a_{0} \} \\
&+a_1 a_2 \{-\Delta_{\perp} +u_{\perp}+u_z+(\Delta_{\perp}-u_{\perp}-u_z)\cos 2a_0 \} \\
&+ 2\theta_p^2 \{ u_{\perp}-u_z+\epsilon_v \cos a_0+ (\Delta_{\perp}-\Delta_z)\cos 2a_0 \} \\
&+2 \epsilon_z (a_1+a_2) \sin a_0. \\
\end{split}
\end{equation}
Because of the linear term, this state is generically unstable in the presence of Zeeman field. For $\epsilon_z=0$, this gives: 
\begin{equation}
\begin{split}
&a_{1,2}: \ u^{H}_{z}=-u^{H}_{\perp}+\Delta_{\perp}-\Delta_z \\
& \ \ \ \ \ \ \ u^{H}_{z}=-u^{H}_{\perp}\pm \sqrt{\Delta^2_z+\epsilon^2_v}+\Delta_{\perp} \\
& \ \ \ \ \ \ \ u^{H}_{z}=-u^{H}_{\perp}-\Delta_z+\Delta_{\perp}\pm \epsilon_v \\
&\theta_p: u^{H}_{z}=\pm \sqrt{\epsilon_v^2+(u^{H}_{\perp}+\Delta_z-\Delta_{\perp})^2}
\end{split}
\end{equation}
\subsubsection{KD-CAF}
The coordinates of the KD-CAF are $a_{1,2}=\cos^{-1} \{\pm \epsilon_z/(\Delta_{\perp}+\Delta_z-u^{H}_{\perp}-u^{H}_z)\}, \ \theta_p=0, \ \theta_s=0, \ \beta=0$. Defining $\cos a_0=-\epsilon_z/(u^{X}_{z}+u^{X}_{\perp})$, the Hartree-Fock functional is : 
\begin{equation}
\begin{split}
E_{HF}&= \frac{1}{2} a^2_1 \{-\Delta_{\perp}+u^{H}_{\perp}+u^{H}_{z}+2(\epsilon_v+ \epsilon_z )\cos a_0+2(-\Delta_{\perp}-2 \Delta_z+ u^{H}_{\perp}+u_z) \cos 2a_{0} \} \\
&+\frac{1}{2} a^2_2\{-\Delta_{\perp}+u^{H}_{\perp}+u^{H}_{z}+2(-\epsilon_v+ \epsilon_z )\cos a_0+2(-\Delta_{\perp}-2 \Delta_z+ u^{H}_{\perp}+u_z) \cos 2a_{0} \} \\
&+a_1 a_2 \{-\Delta_{\perp} +u_{\perp}+u_z+(\Delta_{\perp}-u_{\perp}-u_z)\cos 2a_0 \} \\
&+ \theta_p^2 \{ -\Delta_{\perp}+\Delta_z+u_{\perp}-u_z+(\Delta_{\perp}-\Delta_z-u_{\perp}+u_z) \cos 2 a_0\} \\
&+2 \epsilon_z \cos a_0 \theta_s^2 \\
&+2 \epsilon_v (a_1+a_2) \sin a_0. \\
\end{split}
\end{equation}
Because of the linear term, this state is generically unstable in the presence of hBN substrate. In the absence of it, the linear stability analysis yields: 
\begin{equation}
\begin{split}
&a_{1,2}:\  u^{H}_{z}=-u^{H}_{\perp}+(\Delta_z+\Delta_{\perp}), \\
& \ \ \ \  \ \ \  u^{H}_{z}=-u^{H}_{\perp}+(\Delta_{\perp}-\Delta_{z}), \\
& \ \ \ \  \ \ \  u^{H}_{z}=-u^{H}_{\perp}+(\Delta_z+\Delta_{\perp})\pm \epsilon_z,\\
&\beta:- ,\\
&\theta_s:- ,\\
& \theta_{p}: u^{H}_{z}=u^{H}_{\perp}+(\Delta_z-\Delta_{\perp}).\\
\end{split}
\end{equation}
\subsubsection{SVE}
The instability lines of SVE are:
\begin{equation}
    \begin{split}
       u^{H}_{z}&=-u^{H}_{\perp}+(\Delta_z+\Delta_{\perp})+\epsilon_v-\epsilon_z\\
       u^{H}_{z}&=-u^{H}_{\perp}+(\Delta_{\perp}-\Delta_z)+\epsilon_v-\epsilon_z\\
       u^{H}_{z}&=u^{H}_{\perp}(2 \frac{\Delta_z}{\Delta_{\perp}}-1)+\epsilon_v-\epsilon_z+\Delta_{\perp}+\frac{\Delta_z \epsilon_z}{\Delta_{\perp}}-\Delta_z\\
       u^{H}_{z}&=-u^{H}_{\perp}+\Delta_{\perp}+\epsilon_v-\epsilon_z+\sqrt{\epsilon^2_z-2(2 \Delta_z+\epsilon_v)\epsilon_z+(-2 \Delta_z+\epsilon_v)^2}\\
    \end{split}
\end{equation}
\subsubsection{KD+CAF}
The coexistence phase of KD+CAF of Ref.~\cite{das2022coexistence} can be found by minimizing the Hartree-Fock functional in Eq.~\eqref{S-General HFS} in the subspace of $\theta_{p}=\pi/2$ (since both the KD and the CAF belong in this subspace). More concretely, the HF functional takes the following form then : 
\begin{equation}
\begin{split}
    \mathcal{E}_{HF}= & - \epsilon_z(x_1-x_2)+\frac{1}{4}\Delta_{\perp} \bigg \{(x_1-x_2)^2-(y_1+y_2)^2 \bigg \}+x_1 x_2 \bigg \{-\frac{1}{2}\Delta_z+\frac{3}{2}u^{H}_{\perp}+\frac{1}{2}u^{H}_{z}\bigg \}+y_1 y_2 \bigg \{\frac{1}{2}\Delta_z+\frac{1}{2}u^{H}_{\perp}-\frac{1}{2}u^{H}_{z}\bigg \} \\
    & +\Delta_{\perp}+\frac{\Delta_z}{2}-\frac{u^{H}_{\perp}}{2}-\frac{u^{H}_z}{2}, \\
    \end{split}
\end{equation}
where we have defined $x_{i}(y_i)=\cos a_i \ (\sin a_i), \ i=1,2$, satisfying the constraint $x^2_i+y^2_i=1, \ i=1,2$ . The minimization can be analytically achieved, by changing variables as follows:
\begin{equation}
    \begin{split}
    & z_1=x_1+x_2, \ z_2 =x_1-x_2, \\
    & w_1=y_1+y_2, \ w_2 =y_1-y_2. \\
    \end{split}
\end{equation}
under the constraints :
\begin{equation}
    \begin{split}
     & \sum_{i=1,2} z^2_i+w^2_{i}=4, \\
     & \sum_{i=1,2} z_i w_i=0. \\
     \end{split}
\end{equation}
This yields up to constants:
\begin{equation}
    \mathcal{E}_{HF}=-\epsilon_z w_1 +\frac{\Delta_{\perp}}{2} w^2_1-\frac{u^{H}_{\perp}}{2}u+\frac{w^2_1}{u}(\Delta_z-\Delta_{\perp}-u^{H}_{\perp}-u^{H}_{z}),
\end{equation}
where we have defined $u=w^2_1+w^2_2$. The minimization is straightforward and yields for the coordinates and energy of the coexistence phase KD+CAF, respectively:
\begin{equation}
\begin{split}
&w_1=\frac{-\sqrt{-2 u _{\perp}(\Delta_z-\Delta_{\perp}-u_{\perp}-u_z)}+\epsilon_z}{\Delta_{\perp}} ,\\
&u= \frac{\epsilon_z \sqrt{2(\Delta_z-\Delta_{\perp}-u_{\perp}-u_z)}-2(\Delta_z-\Delta_{\perp}-u_{\perp}-u_z)\sqrt{-u_{\perp}}}{\Delta_{\perp}\sqrt{-u_{\perp}}}, \\
&E_{KD+CAF}=-\frac{1}{2 \Delta_{\perp}}(x-\epsilon_z)^2+C, \ x=\sqrt{-2u_{\perp}(\Delta_z-\Delta_{\perp}-u_{\perp}-u_z)},\\
&C=\Delta_{\perp}+u_{\perp}.
\end{split}
\end{equation}
The concurrence of this state can be easily computed as :
\begin{equation}
    C=\sqrt{(u-w^2_1)w^2_1(4/u-1)},
\end{equation}
and therefore in general the coexistence phase is generally spin-valley entangled. The concurrence is plotted in Fig.~\ref{KD+CAF} for $u^{H}_{z}=-u^{X}_{x}, \ u^{X}_{\perp}=1.25u^{H}_{\perp}, \ \epsilon_z=1$.
\subsubsection{E-KD-AF}
Since the canted phases only in the valleys become unstable in the presence of both fields, we seek to minimize the Hartree-Fock functional in the subspace of $\beta=\theta_s=\theta_p=0$, since all of the states share this feature. Therefore, we find:
\begin{equation}
\begin{split}
    E_{HF}&=2 (\Delta_{\perp}-u^{H}_{\perp})+\cos a_1 \{-2(\epsilon_v+\epsilon_z) +\Delta_z \cos a_1\}+\\
    & 2 \{-\epsilon_v+\epsilon_z+\cos a_1 (-\Delta_{\perp}+u^{H}_{\perp}+u^{H}_{z}) \} \cos a_2+\Delta_z
    \cos^2 a_2 \\
    &= a \cos a_1+\beta \cos^2 a_1+ \gamma \cos a_1 \cos a_2 +\delta \cos^2 a_2+\epsilon \cos a_2. \\
    \end{split}
\end{equation}
We have : 
\begin{equation}
    \begin{split}
     &\frac{\partial E}{\partial a_{1,2}}=0,
      \end{split}
\end{equation}
and the coordinates of this state are given by:
\begin{equation}
    \begin{split}
       a_1 &= \cos^{-1} \{ \frac{\epsilon_z}{\Delta_{\perp}+\Delta_z-u^{H}_{\perp}-u^{H}_z}+\frac{\epsilon_v}{-\Delta_{\perp}+\Delta_z+u^{H}_{\perp}+u^{H}_z}\} ,\\
       a_2 &= \cos^{-1} \{- \frac{\epsilon_z}{\Delta_{\perp}+\Delta_z-u^{H}_{\perp}-u^{H}_z}+\frac{\epsilon_v}{-\Delta_{\perp}+\Delta_z+u^{H}_{\perp}+u^{H}_z}\} .\\
    \end{split}
\end{equation}
We see now that the $\epsilon_{z,v} \rightarrow 0 $ limits reduce to the KD-CAF, CaKD-AF.
\par
\begin{figure}[h!] 
     \centering  
     \includegraphics[width=0.75\textwidth]{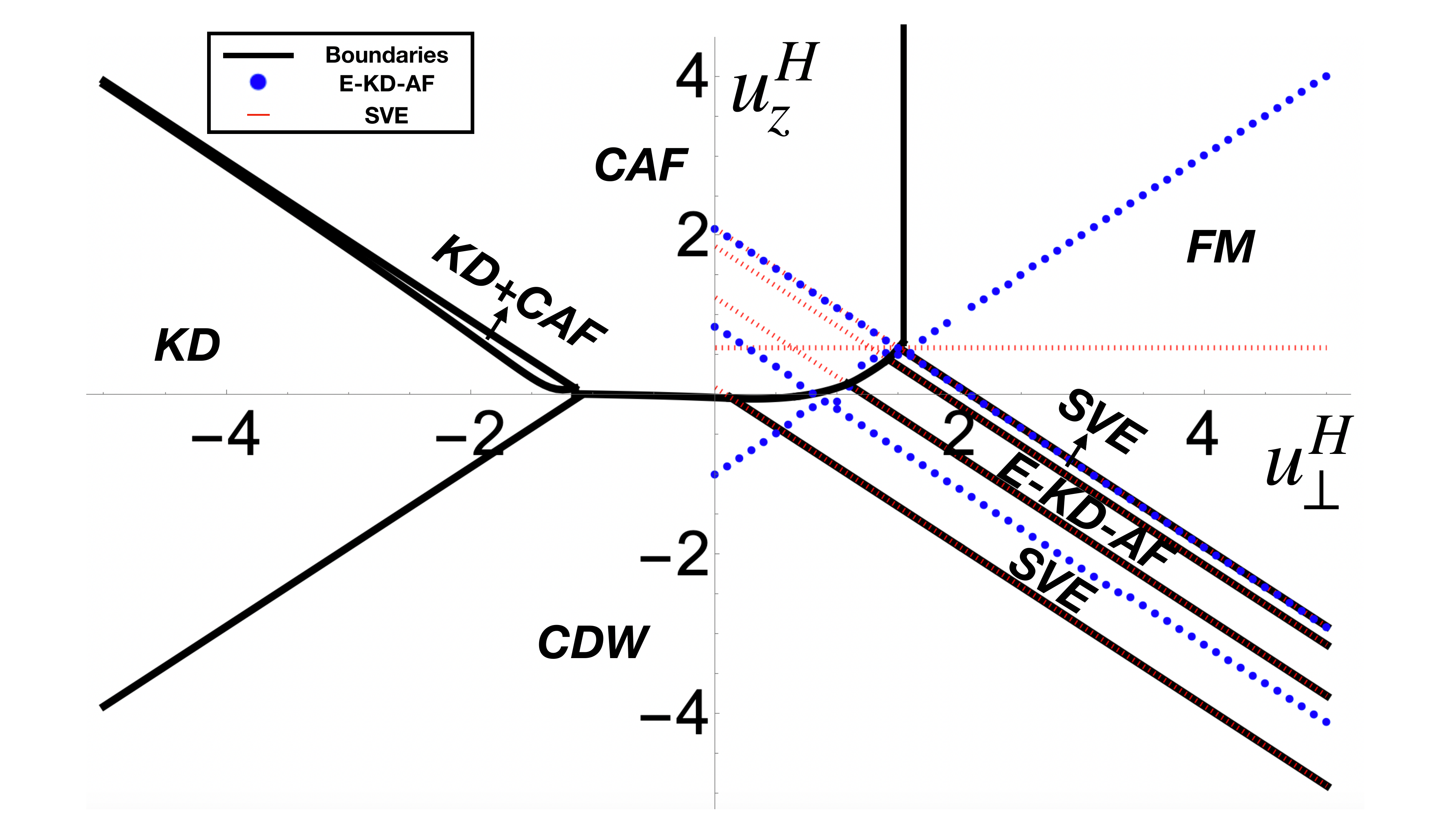}
     \caption{Instability lines of the SVE state and E-KD-AF illustrated as dashed red (blue dotted) for  Zeeman, $\epsilon_z=1$, and valley potential, $\epsilon_v=0.1$, and $\Delta_{\perp}=2, \ \Delta_z=1$. }
     \label{Insta lines}
\end{figure}

 The linearly stable states competing with their energies are summarized in Table~\ref{Table_supp}.
\begin{table}[h!]
\begin{tabular}{|p{2cm}||p{8cm}|} 
 States & Energies\\ [0.5ex] 
 \hline
CDW  & $\Delta_z+u^{H}_{z}-2 \epsilon_v$  \\ 
 \hline
CaKD  &$\Delta_{\perp}+u^{H}_{\perp}- \frac{\epsilon^2_v}{u^{H}_z-u^{H}_{\perp}+(\Delta_z-\Delta_{\perp})} $ \\ 
 \hline
FM  &$2 \Delta_{\perp}+\Delta_z-2u^{H}_{\perp}-u^{H}_{z}-2\epsilon_z$ \\
 \hline
CAF  &$\Delta_z-u^{H}_{z}+\frac{\epsilon^2_z}{2(u^{H}_{\perp}-\Delta_{\perp})}$ \\ 
 \hline
CaKD-AF  &$\Delta_{\perp}-u^{H}_{\perp}-\frac{\epsilon^2_v}{u^{H}_z+u^{H}_{\perp}+(\Delta_z-\Delta_{\perp})} $ \\
\hline
KD-CAF  &$\Delta_{\perp}-u^{H}_{\perp}+\frac{\epsilon^2_z}{ (u^{H}_{\perp}+u^{H}_{z}) -(\Delta_z+\Delta_{\perp})} $ \\[1ex] 
 \hline
E-KD-AF  &$\Delta_{\perp}-u^{H}_{\perp}+\frac{\epsilon^2_z}{u^{H}_{\perp}+u^{H}_z-(\Delta_{\perp}+\Delta_z)}-\frac{\epsilon^2_v}{u^{H}_{\perp}+u^{H}_z+(\Delta_z-\Delta_{\perp})} $ \\
\hline
\end{tabular}
\caption{Energies of linearly stable competing states. The CaKD-AF and KD-CAF remain stable only in the absence of Zeeman effect and hBN substrate respectively.}
\label{Table_supp}
\end{table}
The SVE state has energy : 
\begin{equation}
\begin{split}
E_{SVE}&=-\frac{1}{2\Delta_z}\{(u^{H}_{z})^2+2u^{H}_{z}(\epsilon_z-\epsilon_v+u^{H}_{\perp}-\Delta_{\perp})+\epsilon^2_v-2\epsilon_v(\epsilon_z +u^{H}_{\perp}-\Delta_z-\Delta_{\perp})+\\
&\epsilon^2_z+2 \epsilon_z(u^{H}_{\perp}-\Delta_{\perp}+\Delta_z)+(u^{H}_{\perp}-\Delta_{\perp})^2 
+2\Delta_z(u^{H}_{\perp}-\Delta_{\perp})-\Delta^2_z \} .\\
\end{split}
\end{equation}
The stability lines of the SVE and the E-KD-AF are illustrated in Fig.~\ref{Insta lines}

\section{S-IV: Order parameters of competing states} 
We calculate the order parameters (OPs) of the states $<\hat{O}_{ij}>=<\tau_i \sigma_j>=Tr \{\hat{O}_{ij} \hat{P} \}$. We find that : 
\begin{equation} \label{OP}
    \begin{split}
    <\hat{O}_{ij}>&= \cos^2 \frac{a_1}{2} \braket{\boldsymbol{\eta}|\tau_i|\boldsymbol{\eta}}\braket{\mathbf{s}|\sigma_j|\mathbf{s}}+\sin a_1 \mathbb{R}e \{e^{- i \beta_1}\braket{-\boldsymbol{\eta}|\tau_i|\boldsymbol{\eta}}\braket{-\mathbf{s}|\sigma_j|\mathbf{s}}  \}+ \\
    &+\sin^2 \frac{a_1}{2}\braket{-\boldsymbol{\eta}|\tau_i|-\boldsymbol{\eta}}\braket{-\mathbf{s}|\sigma_j|-\mathbf{s}}+\\
    &+ \cos^2 \frac{a_2}{2} \braket{\boldsymbol{\eta}|\tau_i|\boldsymbol{\eta}}\braket{-\mathbf{s}|\sigma_j|-\mathbf{s}}+\sin a_2 \mathbb{R}e \{e^{- i \beta_2}\braket{-\boldsymbol{\eta}|\tau_i|\boldsymbol{\eta}}\braket{\mathbf{s}|\sigma_j|-\mathbf{s}}  \}+ \\
    &+\sin^2 \frac{a_2}{2}\braket{-\boldsymbol{\eta}|\tau_i|-\boldsymbol{\eta}}\braket{\mathbf{s}|\sigma_j|\mathbf{s}}
    \end{split}
\end{equation}
We note that this reduces for $i \& j \neq 0$ to :
\begin{equation}
    \begin{split}
    <\hat{O}_{ij}>&=\sin a_1 \mathbb{R}e \{e^{- i \beta_1}\braket{-\boldsymbol{\eta}|\tau_i|\boldsymbol{\eta}}\braket{-\mathbf{s}|\sigma_j|\mathbf{s}}  \}+ \\
    &+\sin a_2 \mathbb{R}e \{e^{- i \beta_2}\braket{-\boldsymbol{\eta}|\tau_i|\boldsymbol{\eta}}\braket{\mathbf{s}|\sigma_j|-\mathbf{s}}\}. \\
    \end{split}
\end{equation}
By defining : 
\begin{equation}
    \begin{split}
       \boldsymbol{\eta}(\mathbf{s})= 
       \begin{pmatrix}
       \cos \frac{\theta_{p(s)}}{2}  \\
       \sin \frac{\theta_{p(s)}}{2} e^{i \phi_{p(s)}}  \\ 
       \end{pmatrix} \rightarrow
       -\boldsymbol{\eta}(-\mathbf{s})= 
       \begin{pmatrix}
       \sin \frac{\theta_{p(s)}}{2}  \\
       -\cos \frac{\theta_{p(s)}}{2} e^{i \phi_{p(s)}}  \\ 
       \end{pmatrix},
    \end{split}
\end{equation}
we can calculate the expressions appearing in Eq.~\eqref{OP}.
We find that (we choose $\phi_p=\phi_s=0$) : 
\begin{itemize}
    \item $<\hat{O}_{0i}>=2 s_i (\cos a_1-\cos a_2)$
    \item $<\hat{O}_{i0}>=2 \eta_i (\cos a_1-\cos a_2)$
    \item $<\hat{O}_{xx}>=\eta_z s_z (\sin a_1+\sin a_2)$
    \item $<\hat{O}_{xy}>=<\hat{O}_{yz}>=0$
    \item $<\hat{O}_{xz}>=- \eta_z s_{\perp} (\sin a_1+\sin a_2)$
    \item $<\hat{O}_{yy}>=-\sin a_1+\sin a_2$
    \item $<\hat{O}_{zz}>= \eta_{\perp} s_{\perp} (\sin a_1+\sin a_2)$
\end{itemize}

\subsection{CDW}
The non-vanishing OP is :
\begin{equation}
    <\hat{O}_{z0}>=4 \eta_z \neq 0
\end{equation}
\subsection{CaKD}
The non-vanishing OP is :
\begin{equation}
    <\hat{O}_{i0}>=4 \eta_i \neq 0, \ i=1,2,3
\end{equation}
\subsection{FM}
The non-vanishing OP is :
\begin{equation}
    <\hat{O}_{z0}>=4 s_z \neq 0
\end{equation}
\subsection{CAF}
The non-vanishing OP is :
\begin{equation}
    <\hat{O}_{z0}>=4 s_z \cos a \neq 0
\end{equation}

\subsection{SVE}
The non-vanishing OPs are :
\begin{equation}
    \begin{split}
    <\hat{O}_{0z}>&=2 s_z(1- \cos a) \neq 0 \\
    <\hat{O}_{z0}>&=2 \eta_z(1+\cos a) \neq 0 \\
    <\hat{O}_{xx}>&= <\hat{O}_{yy}>=\sin a \neq 0 \\
    \end{split}
\end{equation}

\subsection{E-KD-AF}
The non-vanishing OPs are :
\begin{equation}
    \begin{split}
    <\hat{O}_{0z}>&=2 s_z(\cos a_1- \cos a_2) \neq 0 \\
    <\hat{O}_{z0}>&=2 \eta_z(\cos a_1+\cos a_2) \neq 0 \\
    <\hat{O}_{xx}>&= \sin a_1 +\sin a_2 \neq 0 \\
    <\hat{O}_{yy}>&=-\sin a_1+\sin a_2 \neq 0 \\
    \end{split}
\end{equation}
\section{S-V: Measure of spin-valley entanglement of states}
A measure of spin valley entanglement is the concurrence, defined as:
\begin{equation}
    C=max\{\lambda_1-\lambda_2-\lambda_3-\lambda_4,0\},
\end{equation}
where $\lambda_i$ are the eigenvalues of the matrix $R= P (\tau_{y} \bigotimes s_{y}) P^{T} (\tau_{y} \bigotimes s_{y}) P$ according to $\lambda_i \geq \lambda_j$, for $i>j$.The matrix $A= (\tau_{y} \bigotimes s_{y}) P^{T} (\tau_{y} \bigotimes s_{y})$ acts as a time-reversal flipping both the valleys and the spin in the density matrix $\rho=\ket{F}_1 \bra{F}_1+\ket{F}_2 \bra{F}_2$. For the valley-active and spin-active states:
\begin{equation}
\begin{split}
    \lambda_1&=\lambda_2=\frac{1}{2} \{ 1-\mathbf{s}_1\mathbf{s}_{2}(\boldsymbol{\tau_{1}}\boldsymbol{\tau_{2}})\},\\
    \lambda_3&=\lambda_4=0,\\
    \end{split}
\end{equation}
where the $s_a$($\tau_a$) matrices are in the case of spin (valley) active states.
\section{S-VI: Comparison with Ref.~\cite{de2023global}}
\label{S-III: Phase diagrams with fixed ratios of the Hartree and Fock parameters}
\par
\begin{figure}[t!] 
     \centering  
     \includegraphics[width=\textwidth]{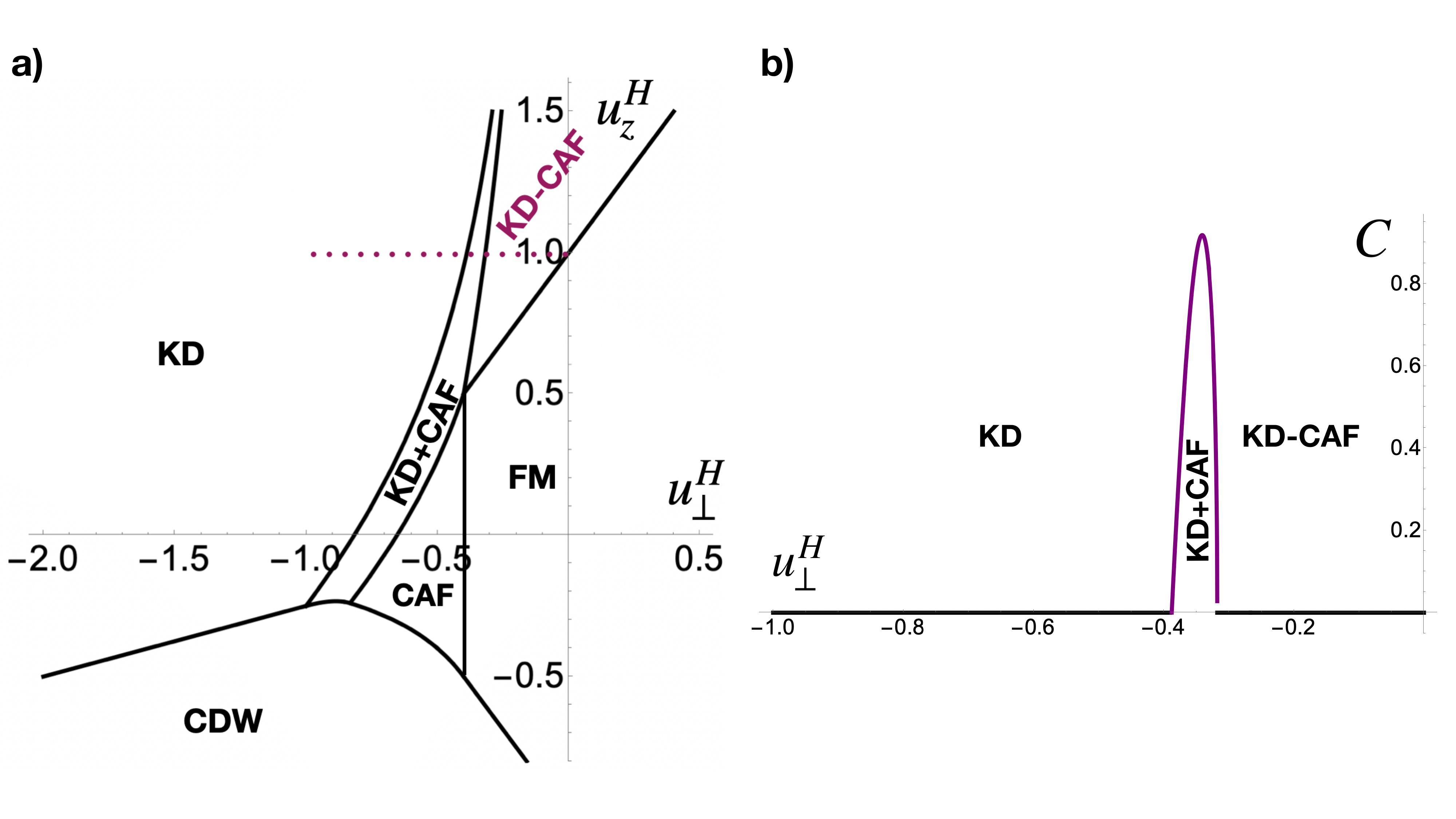}
     \caption{a) Fig.22 of Ref.~\cite{de2023global}. Phase diagram for $u^{H}_{z}=-u^{X}_{x}, \ u^{X}_{\perp}=1.25u^{H}_{\perp}, \ \epsilon_z=1$. b) Concurrence for the parameters of the the phase diagram in a) along the cut $u^{H}_{z}$=1. The concurrence of the KD-CAF vanishes as expected.}
     \label{KD+CAF}
\end{figure}

 In this section we compare in detail some of our results and those of Ref.~\cite{de2023global}. We would like to focus particularly on clarifying the connection between our KD-AF phase and those of Ref.~\cite{de2023global}. The KD-AF state was first identified in Ref.~\cite{stefanidis2023competing}. This state, or more precisely its canted version in spin space (which we call in the main text KD-CAF), was then re-encountered in Ref.~\cite{de2023global} and  labeled as coexistence ``FSVE" phase, where the ``SVE" stands for ``spin-valley entangled". To illustrate this we reproduce Fig.22 of Ref.~\cite{de2023global} in Fig.~\ref{KD+CAF}(a). 

There are, however, two inaccuracies with the identifications of Ref.~\cite{de2023global}. First, while the  KD-CAF state (``FSVE" in Ref.~\cite{de2023global}) has non-trivial spin-valley correlations, it is not a spin-valley entangled state. In fact, in the presence of only spin Zeeman but zero valley potential ($\epsilon_z \neq 0, \ \epsilon_v=0$), from the phases that appear in Fig.22 of Ref.~\cite{de2023global} the only one with non-zero spin-valley entanglement is the coexistence state of AFM and KD (KD+CAF), and all other states, including the KD-CAF (``FSVE" in Ref.~\cite{de2023global}) have zero spin-valley entanglement. This is illustrated in Fig.~\ref{KD+CAF}(b), by plotting the concurrence along the dotted line indicated in Fig.~\ref{KD+CAF}(a).

The second inaccuracy of Ref.~\cite{de2023global} is that the KD-CAF state (``FSVE" in Ref.~\cite{de2023global}), should not be viewed as a coexistence state between two other phases. As we have discussed in the main text, this  KD-AF state simply continuously evolves by canting the spins from the KD-AF state, which is one of the parent states that exist as a phase of its own without the need of single particle Zeeman or valley terms. The KD-AF state is stabilized by the interactions with a range longer than pure delta functions (see Fig.\ref{Fig1}(b) of the main text for $\epsilon_z = 0, \ \epsilon_v=0$). The KD-CAF state contains correlations that combine Kekul\'{e} valley coherence and anti-ferromagnetism, namely, one of the components occupies an equal amplitude superposition of both valleys (e.g. with direction $\boldsymbol{\hat{x}}$ in valley Bloch sphere) with one spin and the other component occupies the opposite valley coherent superposition (e.g. with direction $-\boldsymbol{\hat{x}}$ in valley Bloch sphere) with the opposite spin. Despite combining these characteristics of Kekul\'{e} and AFM state, the KD-AF is however a distinct phase and not merely a coexistence state of these two states (which, for example, would be the case for the coexistence state introduced in Ref.\cite{das2022coexistence}). To see this it suffices to notice that its total XY valley pseudo-spin vanishes (in contrast to Kekul\'{e} state) and so does its staggered sub-lattice spin moment (in contrast to the usual anti-ferromagnet), and therefore it has zero value of the order parameters of these states and thus should not be viewed as a a state where they coexist.


\begin{figure}[t!] 
     \centering  
     \includegraphics[width=\textwidth]{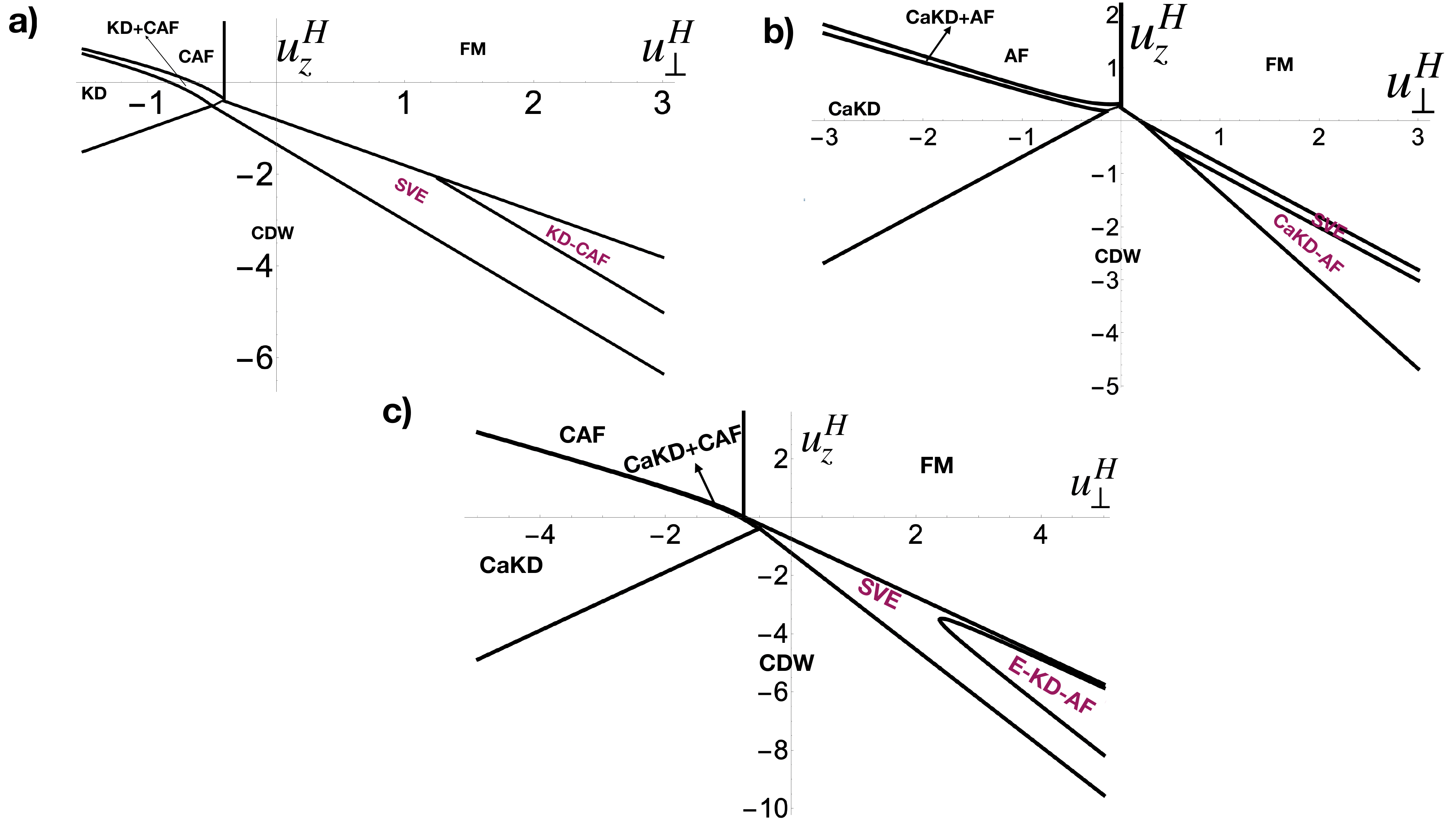}
     \caption{a) Fig. 4 of Ref.~\cite{de2023global}. This is for $\epsilon_z=1, \ u^{X}_{z,\perp}=1.25 u^{H}_{z,\perp}$. There is an additional phase, the KD-CAF, which was missed in Ref.~\cite{de2023global}. b)Fig.14 of Ref.~\cite{de2023global}. This is for $\Delta_{z, \perp}=-0.25 u^{H}_{z, \perp}, \ \epsilon_v=0.25$. There is an additional phase, the CaKD-AF, which was missed in Ref.~\cite{de2023global}. c) Phase diagram when both Zeeman and valley Zeeman fields are present: $\Delta_{z,\perp}=-0.25 u_{z, \perp}, \ \epsilon_v=0.1, \ \epsilon_z=1$}
     \label{Fig_supp}
\end{figure}

We also notice that in several of the phase diagrams of Ref.~\cite{de2023global} the KD-AF and its descendant were missing simply because the range of their plots did not cover the region where this phase becomes favorable. An example of this is illustrated in Fig.~\ref{Fig_supp} which reproduces Fig.4 of Ref.~\cite{de2023global} over a longer paramater range that includes the descendants of the KD-AF state (namely KD-CAF in Fig.~\ref{Fig_supp}(a), CaKD-AF in Fig.~\ref{Fig_supp}(b), E-KD-AF in Fig.~\ref{Fig_supp}(c)).

\end{document}